\documentclass[fleqn,usenatbib]{mnras}

\usepackage{newtxtext,newtxmath}
\usepackage{enumitem}

\usepackage[T1]{fontenc}

\DeclareRobustCommand{\VAN}[3]{#2}
\let\VANthebibliography\thebibliography
\def\thebibliography{\DeclareRobustCommand{\VAN}[3]{##3}\VANthebibliography}

\usepackage{graphicx}	
\usepackage{amsmath}	
\usepackage{color}
\usepackage{ulem}

\title[Short GRB jets in BNS merger environments]{Short gamma-ray burst jet propagation in binary neutron star merger environments}

\author[A. Pavan et al.]{
Andrea Pavan,$^{1,2,3}$\thanks{E-mail: andrea.pavan.20@phd.unipd.it (AP)}
Riccardo Ciolfi,$^{2,3}$\thanks{E-mail: riccardo.ciolfi@inaf.it (RC)}
Jay V. Kalinani$^{1,3}$
and Andrea Mignone$^{4}$
\\
$^{1}$Dipartimento di Fisica e Astronomia, Universit\`a di Padova, Via Francesco Marzolo 8, I-35131 Padova, Italy\\
$^{2}$INAF, Osservatorio Astronomico di Padova, Vicolo dell'Osservatorio 5, I-35122 Padova, Italy\\
$^{3}$INFN, Sezione di Padova, Via Francesco Marzolo 8, I-35131 Padova, Italy\\
$^{4}$Dipartimento di Fisica, Universit\`a di Torino, Via Pietro Giuria 1, I-10125 Torino, Italy
}

\date{Accepted XXX. Received YYY; in original form ZZZ}

\pubyear{2021}

\begin{document}
\label{firstpage}
\pagerange{\pageref{firstpage}--\pageref{lastpage}}
\maketitle

\begin{abstract} 
The multimessenger event GW170817/GRB\,170817A confirmed that binary neutron star (BNS) mergers can produce short gamma-ray burst (SGRB) jets.
This evidence promoted new investigations on the mechanisms through which a BNS merger remnant can launch such a powerful relativistic outflow and on the propagation of the latter across the surrounding post-merger environment.  
In particular, great strides have been made in jet propagation models, establishing connections between the initial jet launching conditions, including the incipient jet launching time (with respect to merger) and the injection parameters,
and the observable SGRB prompt and afterglow emission. 
However, present semi-analytical models and numerical simulations (with one notable exception) adopt simple hand-made prescriptions to account for the post-merger environment, lacking a direct association with any specific merging BNS system. 
Here, we present the first three-dimensional relativistic hydrodynamics simulations of incipient SGRB jets propagating through a post-merger environment that is directly imported from the outcome of a previous general relativistic BNS merger simulation. 
Our results show that the evolution and final properties of the jet can be largely affected by the anisotropies and the deviations from axisymmetry and homologous expansion 
characterizing more realistic BNS merger environments.
In addition, we find that the inclusion of the gravitational pull from the central compact object, often overlooked, can have a major impact. 
Finally, we consider different jet launching times referred to the same BNS merger model and discuss the consequences for the ultimate jet properties. 
\end{abstract}


\section{Introduction}
\label{intro}

The first multimessenger observation of a binary neutron star (BNS) merger in 2017, combining gravitational waves (GWs) with a variety of electromagnetic (EM) signals across the entire spectrum, marked a major milestone in the investigation of these extraordinary astrophysical events (\citealt{LVC-BNS,LVC-Hubble,LVC-MMA,LVC-GRB}; see, e.g., \citealt{Ciolfi2020c,Nakar2020} and refs.~therein).  
Among the numerous discoveries, the coincident detection of the gamma-ray signal GRB\,170817A and the following observation of a multiwavelength afterglow confirmed the long-standing hypothesis that BNS mergers can produce relativistic jets and power short gamma-ray bursts (SGRBs) (\citealt{LVC-GRB,Goldstein2017,Hallinan2017,Savchenko2017,Troja2017,Lazzati2018,Lyman2018,Mooley2018a,Mooley2018b,Ghirlanda2019}).
Moreover, this SGRB was observed from a viewing angle $\approx\!15^\circ-20^\circ$ away from the main jet propagation axis, offering unprecedented insights into the angular structure of the relativistic outflows associated with SGRBs (\citealt{Mooley2018b,Ghirlanda2019}; see, e.g., \citealt{Ioka2019} and refs.~therein).  

Despite this breakthrough discovery, key open questions remain on both the nature of the SGRB central engine (either a massive neutron star or an accreting black hole) and the jet launching mechanism itself.
Observational data from GRB\,170817A and the accompanying afterglow signals directly probed the properties of the relativistic outflow emerging from the baryon-polluted environment surrounding the merger site, but not the physical conditions of the system at the time the incipient jet was initially launched. 

In order to connect the ultimate jet structure and its EM signatures with the properties of the incipient jet (initial opening angle and power, total energy, etc.) and the post-merger environment, including the jet launching time with respect to merger, a growing effort is devoted to model the breakout and propagation of collimated relativistic outflows following BNS mergers. Such an effort, strongly boosted by the observation of GRB\,170817A, includes semi-analytical models (e.g., \citealt{Salafia2020,Lazzati2020,Hamidani2021} and refs.~therein) as well as two- or three-dimensional (magneto)hydrodynamic simulations in the framework of special or general relativity (e.g., \citealt{Nagakura2014, Lazzati2018,Xie2018,Kathirgamaraju2019,Geng2019,Nathanail2020,Murguia2021,Urrutia2021,Nathanail2021,Gottlieb2021} and refs.~therein).

While the physical description provided by the above modelling effort is continuously improving, current studies share one important limitation: 
the density, pressure, and velocity distributions characterizing the surrounding environment at the jet launching time are the result of hand-made prescriptions that should reproduce a typical post-merger system, but have no direct connection with any specific merging BNS.\footnote{An exception is represented by \citet{Nativi2021}, where the environment is set by importing data from a newtonian simulation of neutrino-driven winds produced by a massive neutron star remnant \citep{Perego2014}.}

As a first step towards a consistent end-to-end description covering merger, jet launching, and jet propagation, we present here the first three-dimensional (3D) special relativistic hydrodynamic simulations of incipient SGRB jets where the initial conditions of the surrounding environment are directly imported from the outcome of a fully general relativistic BNS merger simulation.
We discuss the details of the setup and the adopted prescriptions, along with the results of a number of simulations testing different aspects of our approach.
Our findings on a fiducial model reveal the severe limitations of employing hand-made environment initial data as opposed to the outcome of actual BNS merger simulations. Moreover, they demonstrate the importance of including the effects of the gravitational pull from the central object.
Referring to the same merging BNS, we vary the time (after merger) at which the jet is launched and show how this affects the final structure and properties of the escaping jet.
This work serves mostly as a demonstration of the approach, paving the way for future explorations of the relevant parameter space and the first application to events like GRB\,170817A. 

The paper is organized as follows. Section~\ref{setup} presents our setup in terms of numerical methods, initial data, grid structure, boundary conditions, and jet injection properties. Our fiducial model is discussed in Section~\ref{fiducial100}, where we also consider test cases where we remove the contribution of external forces (including gravity) and, in one case, we also  substitute the initial surrounding environment with a much simpler matter distribution inspired by prescriptions typically adopted in the literature.  
In Section~\ref{fiducial200}, we consider a different jet injection time and discuss the impact on the final outcome. Finally, a summary of the work and concluding remarks are given in Section~\ref{summary}.

\section{Physical and numerical setup}
\label{setup}

We perform our special relativistic hydrodynamic simulations using the publicly available code PLUTO, version 4.4 \citep{Mignone2007-PLUTO1,Mignone2012-PLUTO2}. 
The code provides a multi-physics, multi-algorithm modular environment designed to solve conservative problems in different spatial dimensions and systems of coordinates, especially in presence of strong discontinuities. 
We carry out our simulations using the HLL Riemann solver, piecewise parabolic reconstruction and $3^{\rm rd}$-order Runge Kutta time stepping in 3D spherical coordinates $(r,\theta,\phi)$. 
When setting the computational domain in the radial direction, we ``excise'' the central region up to a radius of $r_\mathrm{exc}\!=\!380$\,km, i.e.~we do not evolve the inner part of the system. A careful choice of the boundary conditions on the corresponding spherical surface allows for angle dependent ingoing and outgoing fluxes according to the combined effects of gravitational pull and radial pressure gradients (see Section~\ref{gravity}). Moreover, the incipient jet is introduced into the computational domain from the same surface and thus our jet prescription (Section~\ref{jet}) refers to its properties at 380\,km from the central engine.
We also note that general relativistic effects, which are not accounted for in PLUTO, can be safely neglected above 380\,km. 

The initial setup of our simulations is based on the outcome of a previous general relativistic BNS merger simulation, from which data are imported. 
The physical and numerical setup of such BNS merger simulation is identical to the one employed in \citet{Ciolfi2020a}, except that in this case magnetic fields are not present. 
In particular, the BNS system at hand has the same chirp mass as the one estimated for GW170817 \citep{LVC-170817properties}, with mass ratio $q\!\simeq\!0.9$, and the equation of state (EOS) adopted for neutron star (NS) matter is a piece-wise polytropic approximation of the APR4 EOS \citep{Akmal:1998:1804} as implemented in \cite{Endrizzi2016}. The above choices lead to a long-lived supramassive NS as the merger remnant, which would survive the collapse to a black hole (BH) for much longer than the evolution time covered by the simulation, i.e.~up to 156\,ms after merger.
The BNS merger simulation employs a 3D Cartesian grid with 7 refinements levels and finest grid spacing of $\approx\,$250$\,$m, extending up to $\approx\,$3400$\,$km along all axes. To save computational resources, we also enforced reflection symmetry across the $z\!=\!0$ equatorial plane. An artificial constant density floor of $\rho^*\simeq6.3\times10^4\,$g/cm$^3$ is also set in the numerical domain, corresponding to a total mass of $\simeq3.5\times10^{-3}\,M_{\odot}$.
We refer the reader to \citet{Ciolfi2017,Ciolfi2019} and \citet{Ciolfi2020a} for further details about numerical codes and methods.

In this work, we adopt the paradigm in which a SGRB jet is launched by the accreting BH system resulting from the eventual collapse of the massive NS remnant. While the SGRB central engine is still a matter of debate, this scenario remains the leading one and also finds support in BNS merger simulations (\citealt{Ruiz2016,Ciolfi2020a}; see, e.g., \citealt{Ciolfi2020b} for a review). 
For the time being, we assume that the collapse occurs at a chosen time along the massive NS remnant evolution and we use the physical conditions of the system at that time to start our PLUTO simulations. After a short (order $\sim\!10$\,ms) transition interval to account for the effects of the forming BH-accretion disk system, we introduce the jet with a given set of properties. 
This choice allows us to control the parameters of the injection, which is very convenient for a first investigation, and to explore the effects of a different collapse time on an otherwise identical system.\footnote{Importing data from BNS merger simulations directly covering the collapse and the formation of an incipient jet would represent a further crucial step towards a fully consistent end-to-end description and should be the goal of future studies.}

In the following, we discuss in detail our prescriptions, including data import, grid settings, boundary conditions, treatment of external forces, jet injection, and more.
\begin{figure*}
	\includegraphics[width=1.6\columnwidth]{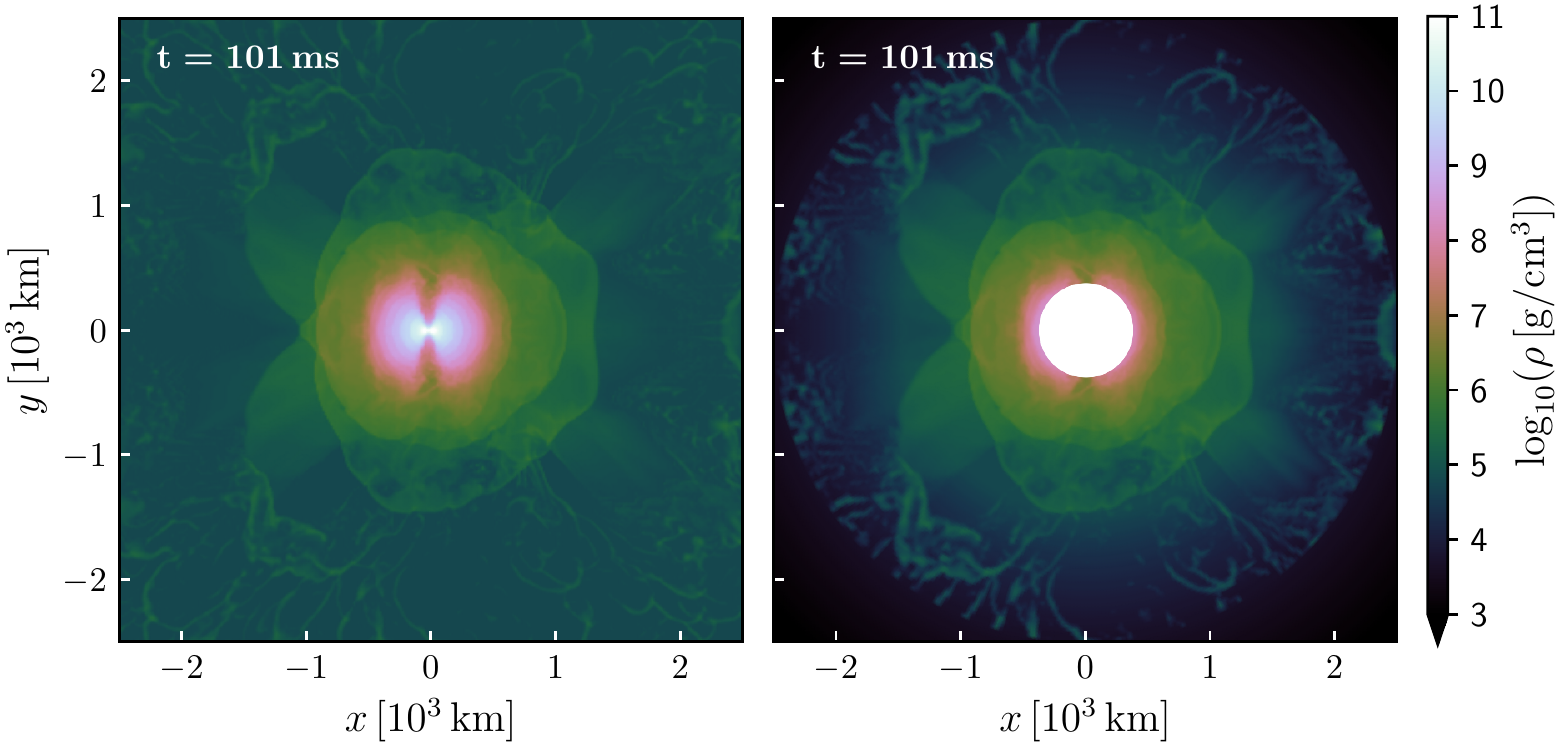}
    \caption{Meridional view of the rest-mass density as obtained by importing data from the reference BNS merger simulation at 101$\,$ms post-merger (see text). 
The left panel shows the result of the PostCactus interpolation, while the right panel shows the final setup on the PLUTO grid. The white circle of 380\,km radius in the right panel corresponds to the excised region (that we do not evolve in PLUTO).}
    \label{fig0}
\end{figure*}

\subsection{Data import and computational grid}
\label{import}

Our reference BNS merger simulation leads to a remnant NS of gravitational mass $M_0\!\simeq\!2.596\,M_{\odot}$ and follows its evolution up to 156\,ms after merger. Along the evolution, we save 3D outputs of rest-mass density, pressure, 3-velocity, and specific internal energy every $\simeq\!5$\,ms. 
For a chosen time at which the remnant NS is assumed to collapse, we import the corresponding data in PLUTO according to the following steps:

\begin{itemize}[leftmargin=+.3cm]
\item First, data are mapped onto a uniform 3D Cartesian grid using the PostCactus Python package\footnote{\url{https://github.com/wokast/PyCactus}} and setting the resolution of the new grid to be the same of the BNS merger simulation at $380\,$km from the origin (i.e.~$\simeq\!8.2$\,km, sixth refinement level).
In this step, we exploit the equatorial symmetry to obtain the data on the full domain (i.e.~for both positive and negative $z$).
\item Second, we apply a $90^{\circ}$ clockwise-rotation around the $x$-axis. In this way, $y\!=\!0$ becomes the new equatorial plane (and reflection symmetry plane) of the BNS merger simulation, while the $y$-axis becomes the new orbital axis. 
This avoids dealing with the singularity at $\theta=0$ in our spherical coordinate system.
\item Then, data are imported in PLUTO and interpolated onto a 3D spherical coordinate grid where we remove the region $r\!<\!r_\mathrm{exc}(=380\,\mathrm{km})$. 
A logarithmic increase in the grid spacing along the radial direction is adopted, allowing us to retain high resolution close to the inner boundary, where it is required, while significantly lowering the number of grid points at larger and larger distances.
\item The artificial density floor or ``atmosphere'' employed in the BNS merger simulation ($\simeq6.3\times10^4\,$g/cm$^3$) is appropriate within a distance of order $\sim\!1000$\,km, but at larger scales it needs to be replaced with a medium with density and pressure that rapidly decrease with distance.
More specifically, we import density and pressure values in the region $r_\mathrm{exc}\le r < 1477$\,km, while at larger radii we replace the artificial floor contribution with a function decaying as $r^{-a}$, where $a\!=\!5$.
At $r\!>\!2500\,$km, we only retain such decaying artificial atmosphere and do not use anymore data imported from the BNS merger simulation. 
In this way, we have the freedom to fill the remaining computational domain up to the outer radial boundary, which is set to $r_{\max} = 2.5\times10^6$\,km. 
In Appendix~\ref{atmo}, we analyze the impact of the atmosphere on the final outcome of our simulations by showing the results obtained with different power-law exponents for the decaying density and pressure.
\end{itemize}

\begin{figure}
	\includegraphics[width=\columnwidth]{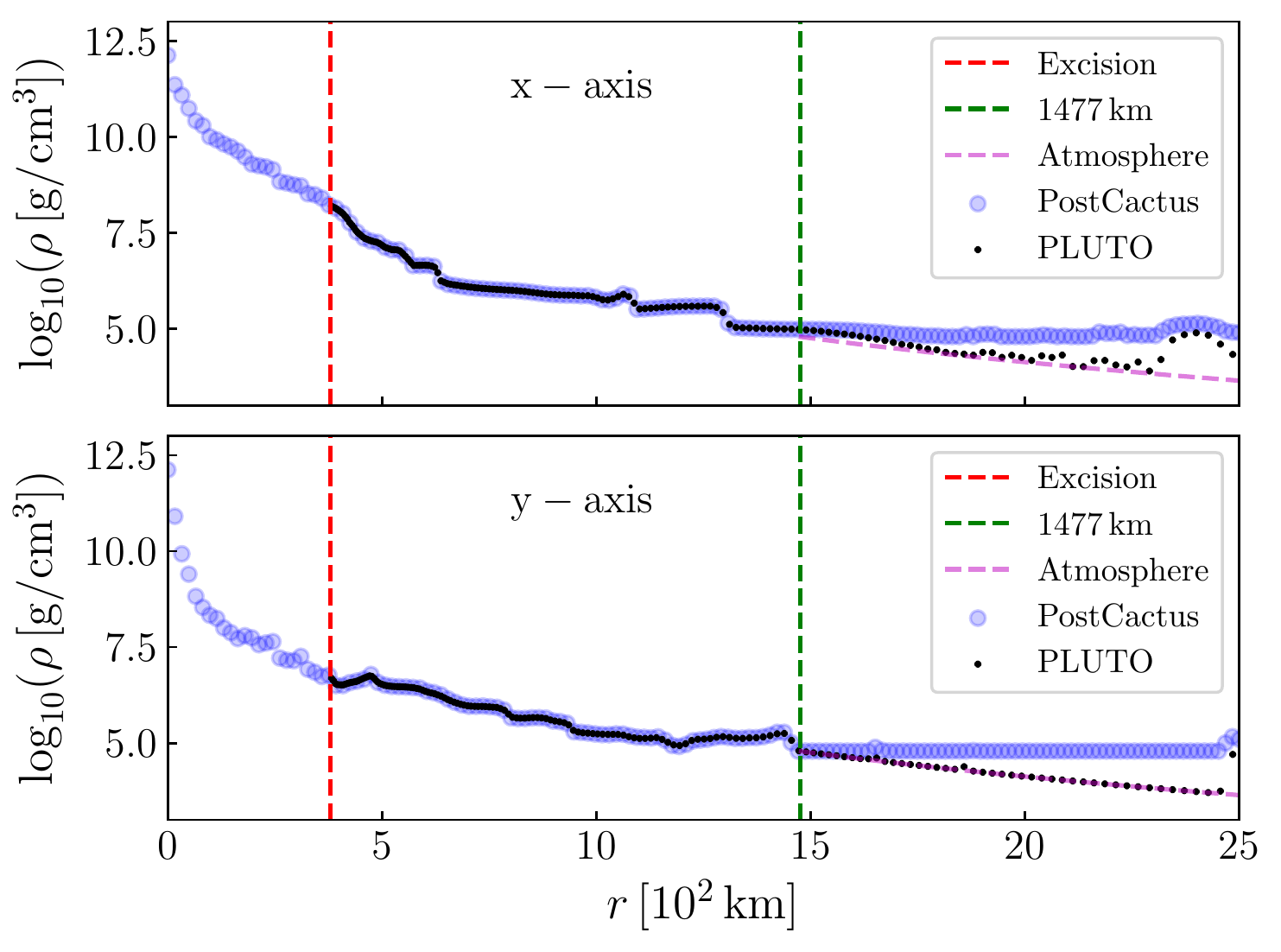}
    \caption{Radial profile of the rest-mass density along the $x$- and $y$-axes for data imported from the reference BNS merger simulation at 101$\,$ms post-merger.
The blue and black dots refer to the result of the PostCactus interpolation and to the final setup in PLUTO, respectively. The vertical red- and green-dashed lines indicate, respectively, the excision radius and the radial distance (1477$\,$km) above which we replace the uniform artificial density floor with a profile decaying as $r^{-5}$ (shown with a magenta-dashed line). See text for further details.}
    \label{import-1D}
\end{figure}

As an example of the PostCactus interpolation, we report in the left panel of Figure~\ref{fig0} the results obtained for the rest-mass density at $101\,$ms after merger. On the right panel of the same Figure, we show instead the final result of the procedure discussed above to import data into the PLUTO computational grid. 
To better illustrate the corresponding atmosphere replacement, we also show in Figure~\ref{import-1D} the radial profiles of the rest-mass density along the $x$- and $y$-axes.

The physical quantities from the BNS merger simulation that are used for setting up the initial data are rest-mass density, pressure, and 3-velocities, while the specific internal energy is recomputed via the EOS.
In PLUTO, we employ the Taub EOS, which corresponds to an ideal gas EOS with $\Gamma=4/3$ in the highly relativistic limit and $\Gamma=5/3$ in the non-relativistic limit, with a smooth and continuous behaviour at intermediate regimes \citep{Mignone2007}.
Since this EOS does not match exactly the EOS of the BNS merger simulation at the low densities of interest, the specific internal energy in the PLUTO setup does not coincide with the one of the original BNS merger data. 
In order to ensure that such a mismatch has no relevant impact on the final conclusions of our study, we performed twice the same simulation where either (i) the pressure is directly imported and the specific internal energy is derived from the Taub EOS or (ii) the opposite. The comparison is discussed in Appendix~\ref{EOS}.

As fiducial resolution, we adopt $756\times252\times504$ points along $r$, $\theta$, and $\phi$, respectively. With a logarithmic radial grid, this yields the smallest grid spacing (at $r_\mathrm{exc}\!=\!380$\,km) of $\Delta r\!\simeq\!4.4$\,km, $r \Delta \theta\!\simeq\!4.4$\,km, and $r \Delta \phi\!\simeq\!4.7$\,km.
We also note that, to avoid the polar axis singularity, $\theta$ varies within the range $[0.1,\pi-0.1]$, while $\phi$ covers the whole $[0,2\pi]$ interval. 
A resolution study is presented in Appendix~\ref{res}, where we show results for the same model with four different resolutions (including the fiducial one).

\subsection{Gravity, pressure gradients, and boundary conditions}
\label{gravity}

While general relativistic effects are not important at $r\!\gtrsim\!380$\,km, the (Newtonian) gravitational pull from the central object remains an ingredient that must be taken into account. 
In particular, gravity causes the fall-back of the inner part of the slowly expanding material that constitutes the surrounding environment through which the incipient jet has to drill and, as we demonstrate in Section~\ref{forces} (see also Figure~\ref{1DExtrap_NEW}), this has an impact on the final jet energetics and collimation.  
For this reason, we introduce the Newtonian gravitational acceleration 
\begin{equation}\label{g}
    \Vec{g} = -\,G\dfrac{M_0}{r^2}\hat{r}\, , 
\end{equation}
where $M_0$ is the gravitational mass of the merger remnant (specified above) and $G$ is the gravitational constant.\footnote{External forces (including gravity) are introduced in our simulations using the BodyForceVector() function provided by PLUTO. In order to consistently treat them in the special relativistic case, we corrected the relevant equations in the latest PLUTO release (version 4.4). More details can be found in the PLUTO User's Guide at \url{http://plutocode.ph.unito.it/documentation.html}.} 

Before introducing the collapse to a BH and the subsequent launching of a jet, we tested our ability to simply reproduce the ongoing evolution of the remnant NS and surrounding environment, based on the information provided by the BNS merger simulation. 
As we show in Appendix~\ref{extr}, a simple and reliable prescription can be adopted once the post-merger dynamics, under the combined action of gravitational pull, centrifugal support, and pressure gradients, has settled to a quasi-stationary state (i.e. later than $\sim\!120$\,ms after merger, for the case at hand).
At this stage, the angle-averaged radial velocity at $r\!=\!380\,$km is nearly constant in time, while the rest-mass density and pressure show an approximately linear increase (Figure~\ref{meandata_126}). The prescription consists of imposing, as radial boundary conditions at the excision radius, the initial rest-mass density, pressure, and 3-velocity as imported from the BNS merger simulation (with their original angular distributions) multiplied by a time-dependent factor that reflects the above angle-averaged trends. 
For $\theta$ and $\phi$ coordinates, we impose instead zero-gradient (i.e.~``outflow'') and periodic boundary conditions, respectively.
A direct comparison with the original evolution up to 156\,ms after merger (final time of the BNS merger simulation) shows a nice match in all quantities (Appendix~\ref{extr}). In contrast, this is no longer the case when gravity is neglected (Figure~\ref{1DExtrap_NEW}).

The above result is particularly relevant as a basis to extrapolate the evolution beyond what is originally covered by the BNS merger simulation. In Section~\ref{fiducial200}, as an example, we exploit it to study the case in which the remnant NS is assumed to collapse at $201$\,ms after merger (i.e. 45\,ms beyond the reach of the original merger simulation).
\begin{figure*}
   \includegraphics[width=2.0\columnwidth,keepaspectratio]{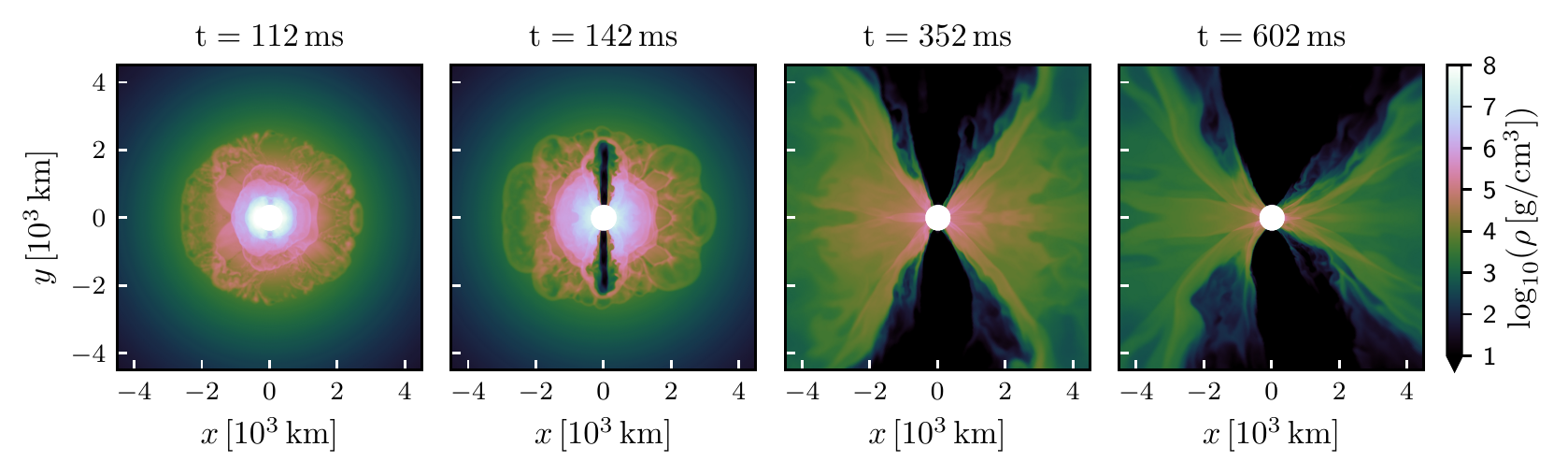}
   \caption{Meridional view of the rest-mass density at different times for our fiducial model (see Section~\ref{fiducial}). Left panel refers to the initial jet launching time. As in Figure~\ref{fig0}, the white circle of 380\,km radius corresponds to the excised region.}
   \label{rho_small}
\end{figure*}

We now turn to consider the adopted prescription to handle the evolution after the remnant NS is assumed to collapse.
In this case, there is no direct information from the BNS merger simulation on how the evolution should proceed. As discussed above (see also Section~\ref{forces}), gravitational pull is a necessary ingredient. We introduce it along with zero-gradient radial boundary conditions, thus allowing the material to eventually cross our inner boundary $r_\mathrm{exc}\!=\!380\,$km while falling-back towards the center. 
Boundary conditions for $\theta$ and $\phi$ coordinates, are again zero-gradient and periodic, respectively.

In addition, there is another aspect that should be taken into account. In a realistic evolution, BH formation does not make the pressure gradient support at 380\,km distance disappear instantaneously, but rather leads to a gradual transition in which such support fades away in time. Moreover, the characteristic timescale for the transition strongly depends on the angle with respect to the spin axis: along this axis, we expect a rather short timescale, no longer than a few tens of ms (see, e.g., \citealt{Ruiz2016}), while orthogonal to it the timescale should be similar to the accretion timescale of the disk surrounding the BH.

In order to introduce such a transition, we add an extra fading-away acceleration term. For simplicity, we consider an isotropic acceleration with the same form of a gravitational force with opposite sign
\begin{equation}\label{push}
    \Vec{a} =G\dfrac{M_\mathrm{eff}(r,t)}{r^2}\hat{r} \, , 
\end{equation}
where the ``effective mass'' $M_\mathrm{eff}$ is a function of both the radial coordinate and time.
At the excision radius and at the time of data import ($t_\mathrm{in}$), we set 
\begin{equation}\label{MHE}
    M_{\mathrm{eff}}(r_\mathrm{exc},t_\mathrm{in})=-\,\left[r^2\dfrac{1}{\Bar{\rho}G}\dfrac{d\Bar{P}}{dr}\right]_{r_\mathrm{exc},t_\mathrm{in}} \, ,
\end{equation}
where $\Bar{P}$ and $\Bar{\rho}$ are the angle-averaged pressure and rest-mass density extracted from the BNS merger simulation. 
This corresponds to having at the inner radial boundary the same initial (angle-averaged) acceleration opposing the gravitational pull as in the original system.
To limit the effect in the vicinity of the excision, we set the radial dependence as a linear decrease such that $M_{\mathrm{eff}}$ becomes zero at a characteristic radius $r^*\!=\!700$\,km (roughly twice the excision radius), i.e. 
\begin{equation}\label{linear}
    M_{\mathrm{eff}}(r,t)=\begin{cases}M_{\mathrm{eff}}(r_\mathrm{exc},t) \frac{r^*-r}{r^*-r_\mathrm{exc}}&\ \mathrm{if}\ r_{\mathrm{exc}}\le r\le r^* \\0&\ \mathrm{if}\ r>r^*\end{cases} \, .
\end{equation}
Finally, we set the time dependence as follows: 
\begin{equation}\label{collapse}
    M_{\mathrm{eff}}(r,t)=M_{\mathrm{eff}}(r,t_c)\exp{\left(-\dfrac{t-t_c}{\tau}\right)} \, ,
\end{equation}
where $t_c$($=\!t_\mathrm{in}$) is the collapse time and $\tau$ is defined as
\begin{equation}\label{timescale}
    \tau \equiv \tau_j-(\tau_d-\tau_j)\sin^2{\alpha} \, ,
\end{equation}
with $\alpha$ the angle with respect to the orbital axis (or the remnant/BH spin axis, i.e.~$\theta,\phi\!=\!\pi/2$). 
In the above expression, $\tau_d$ is the accretion timescale of the BH-disk system, while $\tau_j$ is a characteristic timescale connected to the delay between the collapse and the jet launching time.
Along the BH spin axis, matter is rapidly accreted on a timescale $\tau_j$, allowing the jet to emerge, while on the orbital plane, matter accretes on the much longer timescale $\tau_d$ (see discussion above).
In this work, we set $\tau_j=14.5$\,ms and $\tau_d=0.3$\,s, which is consistent in order of magnitude with what found in BNS merger simulations (e.g., \citealt{Ruiz2016}).

\subsection{Jet injection}
\label{jet}

In our PLUTO simulations, an incipient relativistic jet is introduced into the system from the inner radial boundary ($r_{\mathrm{exc}}=380\,$km), starting $11\,$ms after the time chosen for the collapse of the remnant NS. 
We assume a time-dependent ``top-hat'' (i.e.~uniform) jet contained within a half-opening angle of $10^\circ$ from the $y$-axis, which corresponds to the direction orthogonal to the orbital plane of the BNS merger. 
The injection is two-sided, with identical properties in the $y\!>\!0$ and $y\!<\!0$ regions (we recall that the simulation is in full 3D, without imposed symmetries). 
Outside the injection angle, radial boundary conditions at $r_\mathrm{exc}$ are kept as zero-gradient for the rest-mass density, pressure, and angular components of the 3-velocity ($\mathrm{v}_{\theta},\mathrm{v}_{\phi}$). 
The radial velocity obeys to the same condition as long as $\mathrm{v}_r\!<\!0$, otherwise we set $\mathrm{v}_r=0$. 
Moreover, to ensure numerical stability near the injection region, we change the reconstruction to the more dissipative piecewise linear for $r_{\mathrm{exc}}\le r<385\,\mathrm{km}$ and within an angular distance of $30^{\circ}$ from the $y$-axis. 

For the incipient jet properties at the initial injection time, we set a Lorentz factor $\Gamma_{0}\!=\!3$ (with purely radial outgoing motion), a specific enthalpy $h_0\!=\!100$ (corresponding to a terminal Lorentz factor $\Gamma_{\infty} \equiv h_0\Gamma_0 = 300$), and a two-sided luminosity of
\begin{equation}\label{lum0}
    L_0= 4\pi r_\mathrm{exc}^2\int_0^{\alpha_\mathrm{j}}(h_0\Gamma_{0}^2\rho_0 c^2-P_0) \mathrm{v}_0\sin{\alpha'}d\alpha'=3\times10^{50}\,\mathrm{erg/s}  \, ,
\end{equation}
where $\alpha'$ is the angle with respect to the jet axis, $\alpha_\mathrm{j}$ is the jet half-opening angle in radians, $\mathrm{v}_0$ is the radial velocity (corresponding to $\Gamma_{0}=3$), and $\rho_0$ and $P_0$ are the comoving rest-mass density and pressure, respectively.
In the above expression, $P_0$ is determined from $\rho_0$ and $h_0$ via the Taub EOS. Therefore, $\rho_0$ is the only remaining free parameter and can be adjusted to give the desired $L_0$.

We then impose an exponential time decay in luminosity $L(t) = L_0 \,e^{-t/\tau_d}$, with characteristic timescale $\tau_d=0.3$\,s (the same as the BH-disk accretion timescale; see Section~\ref{gravity}).
Such a decay in luminosity is achieved by means of an exponential decay with double characteristic timescale $2\tau_d$ in both the incipient jet radial velocity $\mathrm{v}_0$ and the term $(h_0\Gamma_{0}^2\rho_0 c^2-P_0)$.

The above incipient jet properties, which are within the expected range for a SGRB jet (e.g., \citealt{Lazzati2020} and refs.~therein), are employed in all the simulations discussed in this work. Investigating the effects of different injection properties is beyond our present scope and will be the subject of future studies.
\begin{figure*}
   \includegraphics[width=2\columnwidth,keepaspectratio]{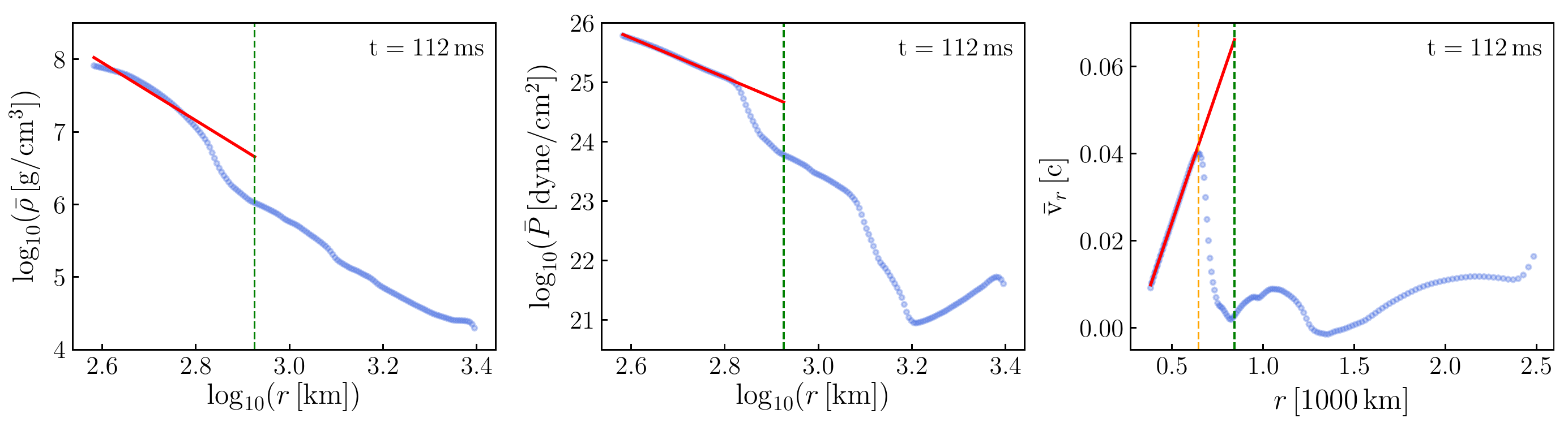}
   \caption{Radial profiles of the angle-averaged rest-mass density, pressure, and radial velocity at 112$\,$ms after merger in our fiducial simulation (blue dots). Red lines represent the analytical fits described in Section~\ref{ball}. The vertical orange dashed line in the right panel marks the distance up to which the nearly homologous expansion regime holds. In all panels, the vertical green dashed line marks the distance at which the analytic density profile gives the same total mass of the environment as the original data.}
   \label{fitbolla}
\end{figure*}

\section{Collapse at 0.1\,seconds after merger}
\label{fiducial100}

In this Section, we discuss the outcome of simulations where we set the collapse time of the remnant NS to $101$\,ms after merger. 
The evolution is followed up to slightly more than 1\,s after merger. 
We start by discussing our fiducial model, where the prescriptions presented in Section~\ref{setup} are applied in full. 
In addition, we consider two more simulations, one where we do not include the external forces (i.e.~the gravitational pull and the extra outward acceleration accounting for the fading-away radial pressure gradients close to the excision; see Section~\ref{gravity}) and another one where we additionally replace the post-merger environment with a spherically symmetric matter distribution in homologous expansion, as often assumed in SGRB jet propagation models.

\subsection{Fiducial model}
\label{fiducial}
For our first jet simulation in PLUTO (hereafter ``fiducial'' case or model), we started from BNS merger data imported at 101\,ms post-merger as described in Section~\ref{import}. The corresponding initial data for the rest-mass density are shown in the right panel of Figure~\ref{fig0}.
The excised region (of radius 380\,km) is surrounded by a slowly expanding (maximum radial velocity $\simeq\!0.07\,c$) cloud of material of mass $\simeq\!0.02\,M_\odot$ and extending up to a radius of $\sim\!2000$\,km, with density declining with distance by a few orders of magnitude. The higher density inner region ($380\,\mathrm{km}\!\leq\!r\!\lesssim\!500$\,km) presents a significant deviation from an isotropic distribution, with a lower density funnel along the orbital axis (or remnant NS spin axis). 
This is a common feature observed in BNS merger simulations (e.g., \citealt{Ciolfi2020b} and refs.~therein), resulting from the combination of the gravitational pull and the non-isotropic centrifugal support and pressure gradients.
As we discuss in this Section, the presence of such a funnel can significantly affect the initial propagation of an incipient jet. 

Following the prescriptions described in the previous Section~\ref{setup}, we assume that 101\,ms is the time at which the remnant NS collapses to a BH.
After 11\,ms of evolution accounting for the formation of a central BH-disk system, i.e.~at 112\,ms after merger, we inject into the system a relativistic beam with the chosen properties (specified in Section~\ref{jet}).
In the first panel of Figure~\ref{rho_small}, we show the rest-mass density distribution at the initial time of injection.
Due to the further outflow of matter emerging from the excision surface, the total mass of the surrounding environment is about 30\% larger with respect to 101\,ms. 
In Figure~\ref{fitbolla}, we report the angle-averaged rest-mass density, pressure and radial velocity at the same time.
From the right panel, we notice that the expansion is nearly homologous up to more than 600\,km.

As the incipient jet starts to propagate though the surrounding environment, a high collimation is maintained up to the breakout time, around 30\,ms later (Figure~\ref{rho_small}, second panel).
As the injection continues, more and more energy is transferred laterally to the material surrounding the jet, leading to a hot and high-pressure interface (or cocoon) and eventually to an emerging jet with a certain angular structure (see below).
On small scales, the ensuing evolution is characterized by a widening of the low density funnel excavated by the jet (Figure~\ref{rho_small}, third and fourth panels), due to a changing pressure balance with the surrounding material. 
An important contribution to this effect is given by the continuous accretion of the innermost and most dense material. At 200\,ms after merger, for instance, the mass outside the excised region has already decreased by a factor of $\simeq\!4$.

\begin{figure}
   \includegraphics[width=\columnwidth]{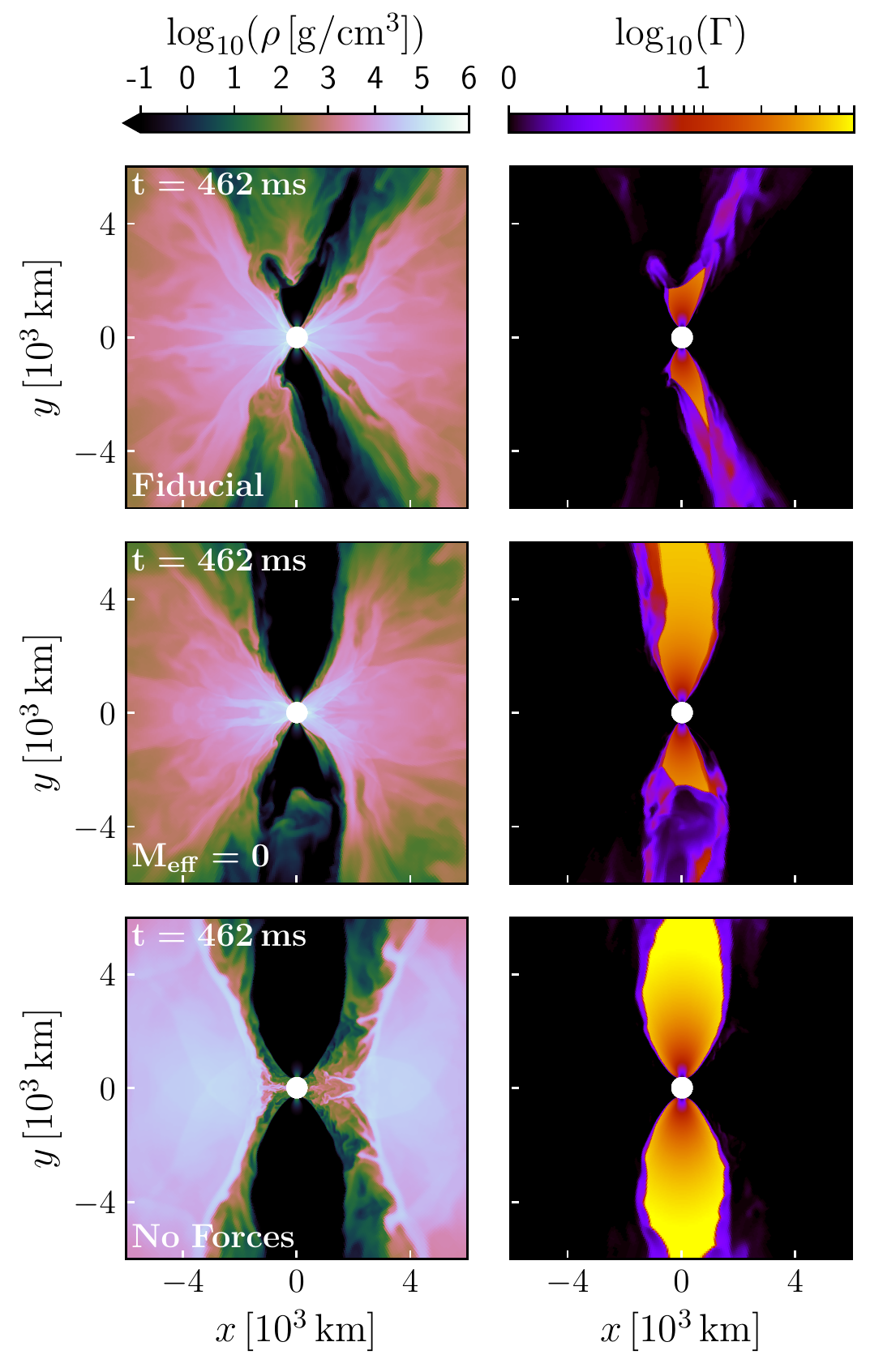}
   \caption{Meridional view of rest-mass density (left) and Lorentz factor (right) at 462\,ms after merger. Top, central, and bottom rows refer to the fiducial simulation (Sect.~\ref{fiducial}), the one with $M_\mathrm{eff}\!=\!0$ (see Sect.~\ref{gravity}), and the one without external forces (Sect.~\ref{forces}), respectively.}
   \label{KH_NEW}
\end{figure}

Along with the above widening effect, the evolution on small scales also reveals the development of Kelvin-Helmholtz instability vortices at the jet-cocoon interface, which lead to episodes of increased baryon loading of the funnel due to portions of material that are occasionally brought in (see, e.g., \citealt{Gottlieb2019} for a discussion of a similar process).
One of this episodes is illustrated in Figure~\ref{KH_NEW} (top panels), where we show a meridional view of rest-mass density and Lorentz factor at 462\,ms after merger. 
Such a process perturbs the recollimation shock at the base of the jet, contributing to the development of intermittency and deviations from axisymmetry, and reducing both the collimation and the overall efficiency in converting the injected power into radial kinetic energy.

In the central panels of Figure~\ref{KH_NEW}, we show the analogous case where we set $M_\mathrm{eff}\!=\!0$, i.e.~we remove the extra outward acceleration term mimicking the fading-away support via radial pressure gradients near the excision. In this case, matter falls back towards the center more rapidly and at 462\,ms after merger the mass above 380\,km radius is already 40\% lower than in the fiducial case. 
The jet-cocoon interface is less turbulent, with only minor episodes of baryon loading within the funnel.
These differences show that a more realistic description of the post-collapse phase near the excision has a potentially relevant impact.

\begin{figure*}
   \includegraphics[height=0.95\textheight]{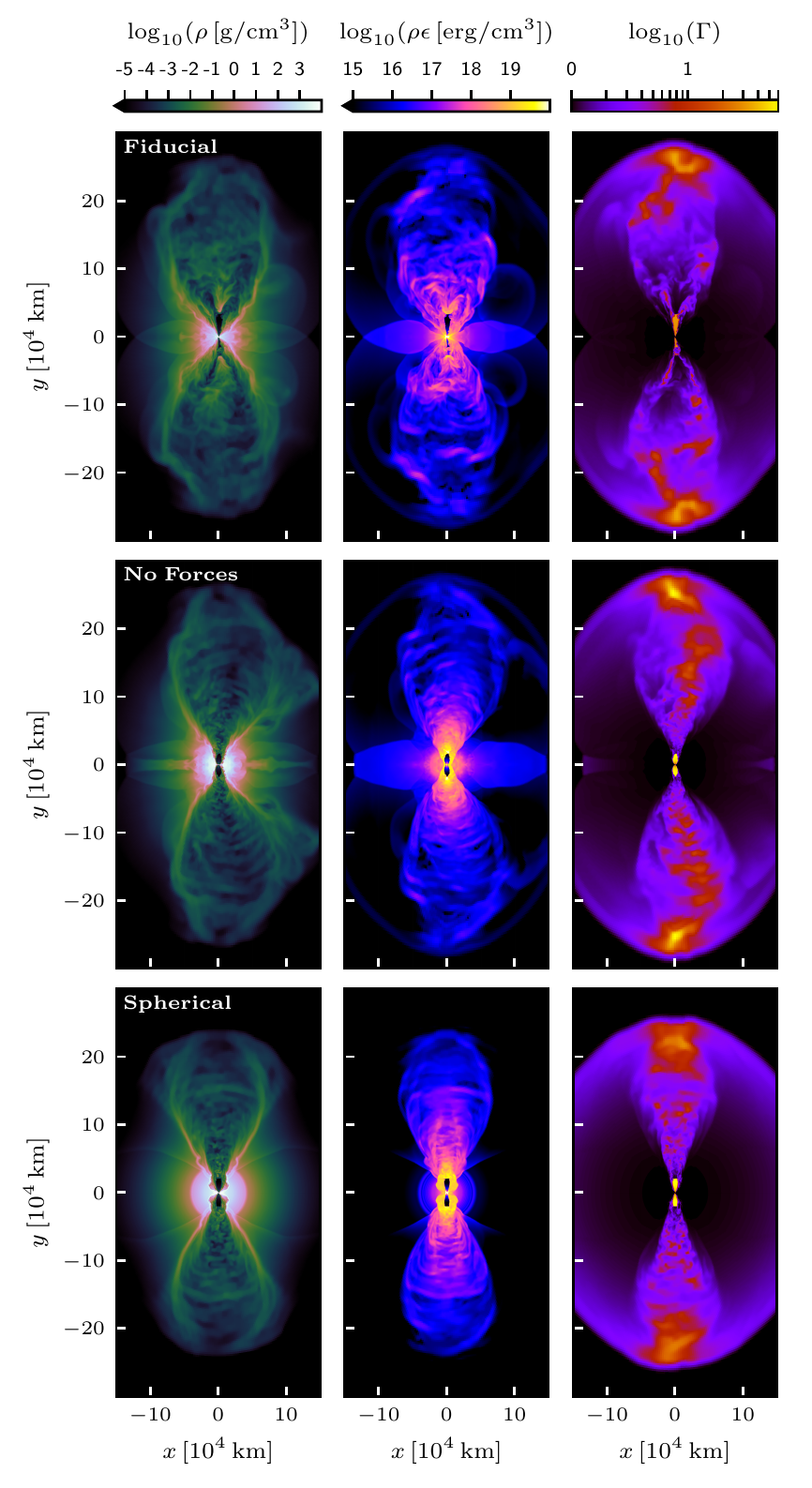}
   \caption{Meridional view of rest-mass density, internal energy density, and Lorentz factor at 1012$\,$ms after merger (left to right). Top, central, and bottom rows refer to the fiducial simulation (Sect.~\ref{fiducial}), the one without external forces (Sect.~\ref{forces}), and the one with simplified isotropic environment (Sect.~\ref{ball}), respectively.}
   \label{large_scale}
\end{figure*}

Figure~\ref{large_scale} (upper panels) shows in full scale (order $\sim\!10^5$\,km) the rest-mass density, internal energy density, and Lorentz factor at the end of our fiducial simulation, i.e.~1012\,ms after merger. At this time, the injection power has significantly declined ($L(t)\propto e^{-t/\tau_d}$ with $\tau_d\!=\!0.3$\,s) and the jet is composed by an ultra-relativistic ``head'' (hereafter referring to the outer high Lorentz factor portion of the outflow) whose front has reached $\simeq\!2.7\times10^5$\,km, followed by a less collimated, slower, hotter, and more turbulent tail.
The maximum Lorentz factor at the jet's head is $\Gamma\!\simeq\!40$.
From the meridional view of the Lorentz factor (Figure~\ref{large_scale}, top right panel), we also notice clear deviations from axisymmetry. 
This reflects the fact that the surrounding environment imported from the BNS merger simulation is not perfectly axisymmetric and the following evolution amplifies further such deviations.
\begin{figure}
   \includegraphics[width=\columnwidth]{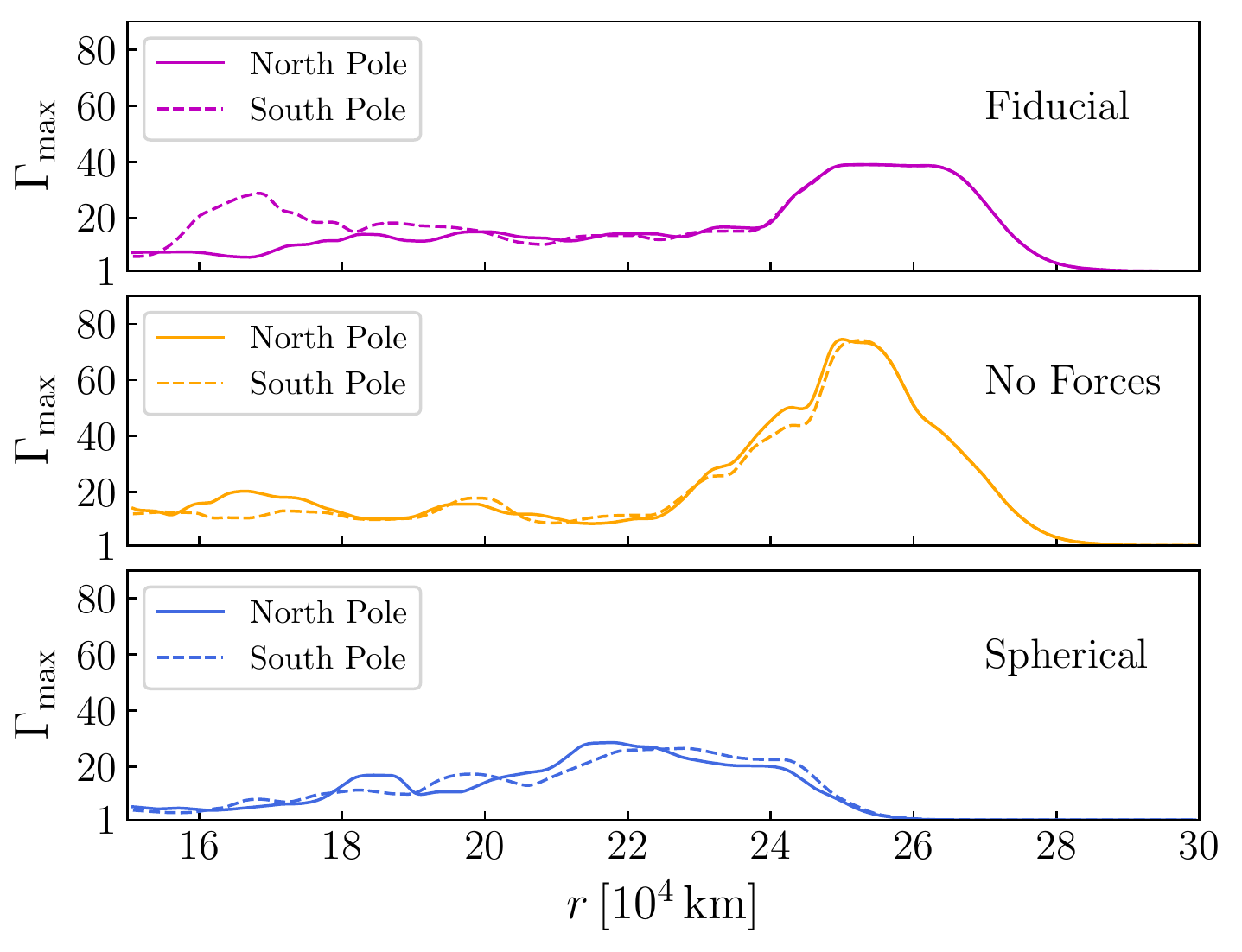}
   \caption{Radial profiles of maximum Lorentz factor at 1012\,ms after merger (maximum value achieved at each radial distance). North and south profiles are shown for the same three cases of Figure~\ref{large_scale}: our fiducial simulation (top), the one without external forces (center), and the one with simplified isotropic surrounding environment (bottom).}
   \label{gamma_max}
\end{figure}

Figure~\ref{gamma_max} (top panel) shows the radial profile of the Lorenz factor at 1012\,ms after merger, where the maximum value is reported for each radius. At the jet's head, north and south profiles are nearly identical to each other. 
We also investigate the angular dependence of the Lorentz factor at the jet's head (and at 1012\,ms after merger), by computing the radial-average $\Bar{\Gamma}$ within the interval $r\!\in (24,27)\times10^4$\,km along different directions. 
In the top-left panels of Figures~\ref{GammaVSalpha_xy} and \ref{GammaVSalpha_yz}, we consider in particular the resulting angular profiles on the $xy$- and $yz$-planes, respectively, referring to the north side only (south profiles are very similar). The profiles are given in terms of the angle $\alpha$, measuring the angular distance from the injection axis (or $y$-axis) and with positive/negative sign for positive/negative $x$ and $z$, respectively. 
On the $xy$-plane, the angular distribution of $\Bar{\Gamma}$ appears very asymmetric between positive and negative $\alpha$ values and the maximum occurs $\simeq\!0.7^{\circ}$ away from the injection axis. On the $yz$-plane, the distribution is only slightly asymmetric and the (higher) maximum is achieved for $\alpha\!\simeq\!0.8^{\circ}$. The rather different profiles in the two planes confirm that the full 3D distribution strongly deviates from axisymmetry. 
The direction containing the absolute maximum of $\Bar{\Gamma}$ in 3D is tilted by $\simeq\!0.9^{\circ}$ with respect to the $y$-axis (or injection axis). 

It is worth noting that angular profiles like those obtained on the $xy$- and $yz$-planes cannot be reproduced by a simple Gaussian or power-law function. In particular, the central peak ($\Bar{\Gamma}\!\gtrsim\!20$) can be nicely fit with a skewed function
\begin{equation}
    \mathrm{S}(\alpha)\propto e^{-\frac{(\alpha-\Bar{\alpha})^2}{2\sigma^2}}\left[1+\mathrm{erf}\left(\beta\dfrac{(\alpha-\Bar{\alpha})^2}{2\sigma^2}\right)\right] \, ,
\end{equation}
with characteristic half-width $\sigma\!=\!1.66^{\circ}$ and $3.34^{\circ}$, respectively (analogous to the Gaussian $\sigma$ parameter).
However, the full profiles present additional lateral wings, particularly prominent on the $xy$-plane, that are hard to fit with any simple function and cannot be neglected when considering the jet energetics (see below).

We also analyze the energy content of the emerging outflow (in the region $r\!>\!3000$\,km) at 1012\,ms after merger. The total kinetic energy is $\simeq\!3.7\!\times\!10^{49}$\,erg, while the internal energy is $\simeq\!5.4\!\times\!10^{48}$\,erg, i.e.~about 15\% of the kinetic one. This ratio confirms a substantial (but not yet complete) conversion of heat into outflowing motion. 
The sum of the above kinetic and internal energies accounts for $\simeq\!48.5\%$ of the total injected energy ($\simeq\!8.7\!\times\!10^{49}$\,erg), where about half of the latter is instead lost due to the gravitational pull acting on the environment material (see also the next Section, where the effects of removing the gravitational pull are discussed).

At the jet's head, taking as a reference the shell given by the radial interval $r\!\in (24,27)\times10^4$\,km and defining the ``core'' as the region within an angle of $\sigma\!=\!3.34^{\circ}$ from the maximum Lorentz factor direction (chosen as the largest $\sigma$ among the $\Bar{\Gamma}$ angular profiles on the $xy$- and $yz$-planes), we obtain  
\begin{align*}
    &\mathrm{E_{kin,core} \simeq 2.947\times10^{48}\,} \mathrm{erg} \, , \  &\mathrm{E_{kin,shell} \simeq 6.806\times10^{48}\,} \mathrm{erg} \, ,\\
    &\mathrm{E_{tot,core} \simeq 3.790\times10^{48}\,} \mathrm{erg} \, , \ &\mathrm{E_{tot,shell} \simeq 8.842\times10^{48}\,} \mathrm{erg} \, .
\end{align*}
In the core, kinetic energy contributes to 78\% of the total energy. The contribution of the core total energy compared to the whole shell is $\simeq 43\%$.

The bottom-left panels of Figures~\ref{GammaVSalpha_xy} and \ref{GammaVSalpha_yz} show the angular profiles (on the $xy$- and $yz$-planes, respectively) of the isotropic equivalent energy $E_\mathrm{iso}$ (kinetic plus internal) of the jet's head, i.e.~within the radial interval $r\!\in (24,27)\times10^4$\,km. Also in this case, only the profiles on the north side are reported, since the ones on the south side are nearly coincident.
As for the radial-averaged Lorentz factor, the $E_\mathrm{iso}$ angular profiles are characterized by a slightly offset and asymmetric central peak that is well reproduced by a skewed function (with characteristic half-width $\sigma\!\simeq\!2^{\circ}$) and by additional lateral wings that are highly (moderately) prominent and asymmetric on the $xy$-plane ($yz$-plane). 
In 3D, $E_\mathrm{iso}$ reaches a maximum of $\simeq\!2.8\times10^{51}$\,erg, occurring about $0.8^{\circ}$ away from the injection axis.

\subsection{Impact of external forces}
\label{forces}

Our second simulation is analogous to the fiducial one (discussed in the previous Section), except that in this case we switch off the acceleration terms that account for the gravitational pull and the fading-away radial pressure gradient support near the excision (see Section~\ref{gravity}).
Although the inclusion of the above external forces makes the simulations arguably more consistent, the corresponding effects are commonly neglected in SGRB jet propagation studies (excluding those based on general relativistic simulations, e.g.~\citealt{Kathirgamaraju2019,Nathanail2021}). 
Here, we aim at assessing whether this choice might have a significant impact on the final jet properties.

A visual comparison at the final simulation time (1012\,ms after merger) is provided in Figure~\ref{large_scale}, where the upper row refers to the fiducial model, while the central row refers to the simulation without external forces. 
Looking at the rest-mass density close to the excision surface, we note that in the latter case a more massive environment surrounds the low density funnel, due to the fact that in absence of a gravitational pull material does not fall back towards the center nor gets accreted. 
As a consequence, not only the widening effect depicted in Figure~\ref{rho_small} is substantially reduced, maintaining a higher degree of collimation,
but also the generation of Kelvin-Helmholtz instabilities discussed in the previous Section is strongly inhibited. To illustrate this, in the bottom panels of Figure~\ref{KH_NEW} we show the rest-mass density and Lorentz factor on the meridional plane at 462\,ms after merger, to be compared with the fiducial case in the top panels of the same Figure.
The shear at the jet-cocoon interface is much more stable and no episodic baryon loading of the funnel is noticed. This results in an essentially unperturbed recollimation shock, which allows the incipient jet to preserve a higher and more stable Lorentz factor.
At the same time, the internal energy density (see Figure~\ref{large_scale}) reveals a hotter and more uniform outflow up to $\sim\!10^5$\,km.

At the end of the simulation (see Figure~\ref{large_scale}), the jet reaches a maximum Lorentz factor almost a factor of 2 higher than the fiducial case, as also reported in Figure~\ref{gamma_max}. The acceleration is however more gradual and the distance reached by the outflow at the final time is similar (if not even slightly smaller). Moreover, because of the higher and more persistent collimation and Lorentz factor at the base of the outflow, the jet's head is now followed by a rather well-defined tail extending down to the excision surface.

Computing the internal and kinetic energies of the emerging outflow for $r\!>\!3000$\,km and at 1012\,ms after merger, we find that they sum up to $\simeq\!1.1\times10^{50}$\,erg, i.e.~about 127\% of the total injected energy. On the one hand, this indicates that the absence of the gravitational pull, unlike the fiducial case, preserves the injected energy. On the other hand, it shows that additional energy residing in the initial environment is carried along by the incipient jet.

From this example, we conclude that neglecting the external forces (in particular gravity) can have major effects on the final outcome.

\subsection{Realistic versus simplified environment}
\label{ball}

For the simulation discussed in this Section, we further reduce the degree of realism by not only neglecting the external forces (like in the previous Section), but also substituting the matter distribution and velocities of the surrounding environment imported from the BNS merger simulation with simpler analytic prescriptions.
In particular, we adopt the common assumptions of (i) a spherically symmetric matter distribution with density and pressure decreasing with radius as a power-law and (ii) homologous expansion. 

In order to produce an initial setup with the above assumptions that is the closest to what we have in the case of imported BNS merger data, we fit the angle-averaged rest-mass density and pressure at 112\,ms after merger with power-law radial functions. Similarly, we fit the angle-averaged radial velocity with a linear function of the radius. 
We limit the fits within a distance of $\simeq\!645$\,km, up to which the three angle-averaged profiles are rather well reproduced (Figure~\ref{fitbolla}). The resulting analytical functions are: 
\begin{align}
  &\Bar{\rho}_{\mathrm{fit}}(r) = 1.056\times10^8\times\left(\dfrac{r}{380\,\mathrm{km}}\right)^{-3.981}\, \mathrm{g/cm^3}\\
  &\Bar{P}_{\mathrm{fit}}(r) = 6.408\times10^{25}\times\left(\dfrac{r}{380\,\mathrm{km}}\right)^{-3.320}\, \mathrm{dyne/cm^2}\\
  &\Bar{\mathrm{v}}_\mathrm{fit}(r) / c = 0.047\times\left(\dfrac{r}{380\,\mathrm{km}}\right)-0.037  \, .
\end{align}
Then, for our simulation setup, we impose an isotropic environment following the above profiles and extended up to a radial distance of $\simeq\!843\,$km. Such a distance is chosen in order to have a total mass of the environment equal to the one in the fiducial simulation. Finally, we add an artificial atmosphere identical to the one imposed in the fiducial case (see Section~\ref{import}).
\begin{figure}
   \includegraphics[width=\columnwidth]{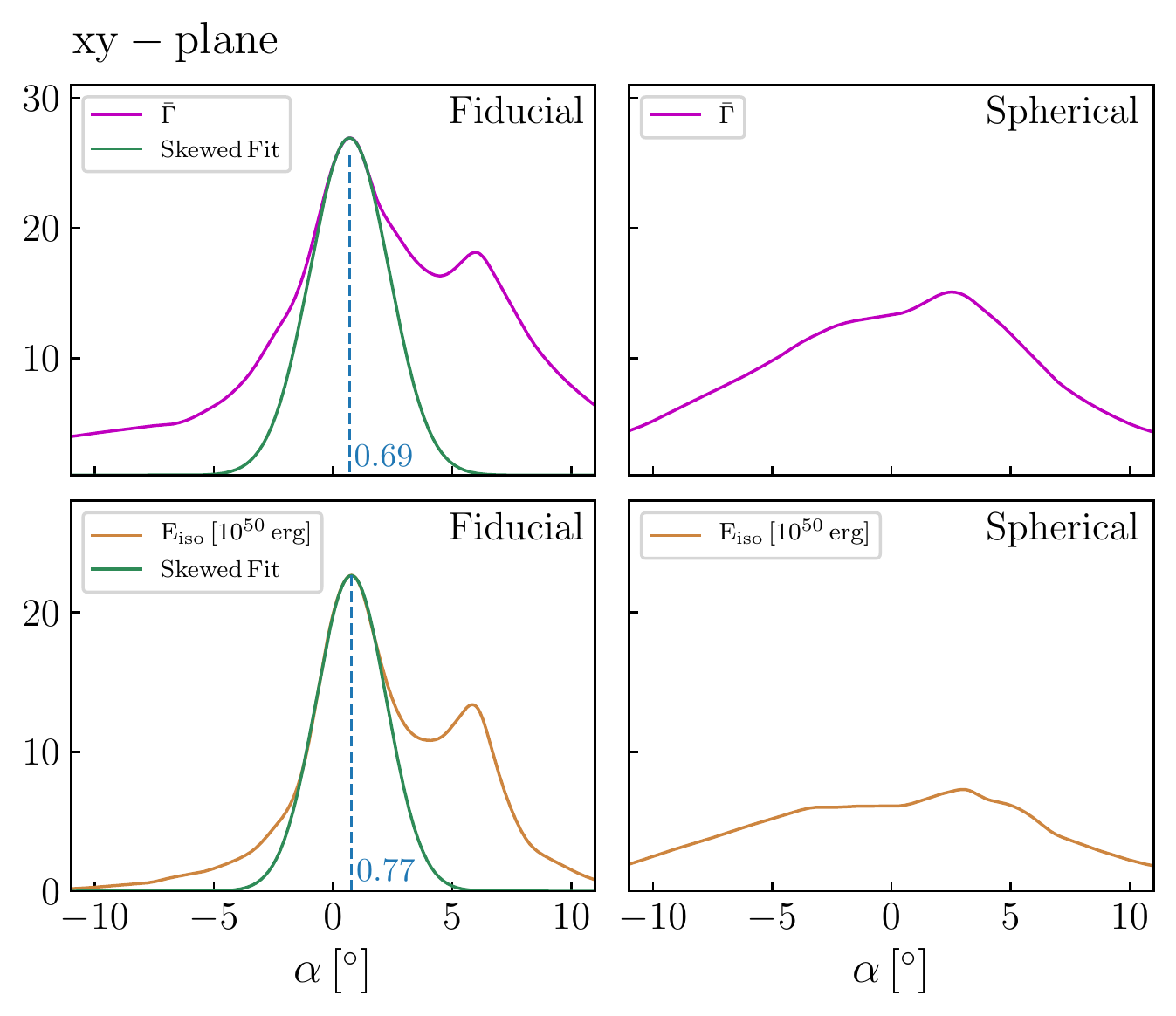}
   \caption{Angular profiles (north side only) on the $xy$-plane of the radial-averaged Lorentz factor (top) and isotropic-equivalent energy $E_\mathrm{iso}$ (bottom) at the jet's head (see text for details) and at 1012\,ms after merger. Left panels refer to our fiducial simulation, while right panels refer to the one with simplified isotropic surrounding environment (Section~\ref{ball}).
Here, $\alpha$ is the angle with respect to the injection axis (i.e. the $y$-axis), with positive/negative sign for positive/negative $x$. 
The green curve in the left panels corresponds to the skewed function that best-fits the central peak of the profile. The vertical dashed blue line marks the angular position of the peak.}
   \label{GammaVSalpha_xy}
\end{figure}
\begin{figure}
   \includegraphics[width=\columnwidth]{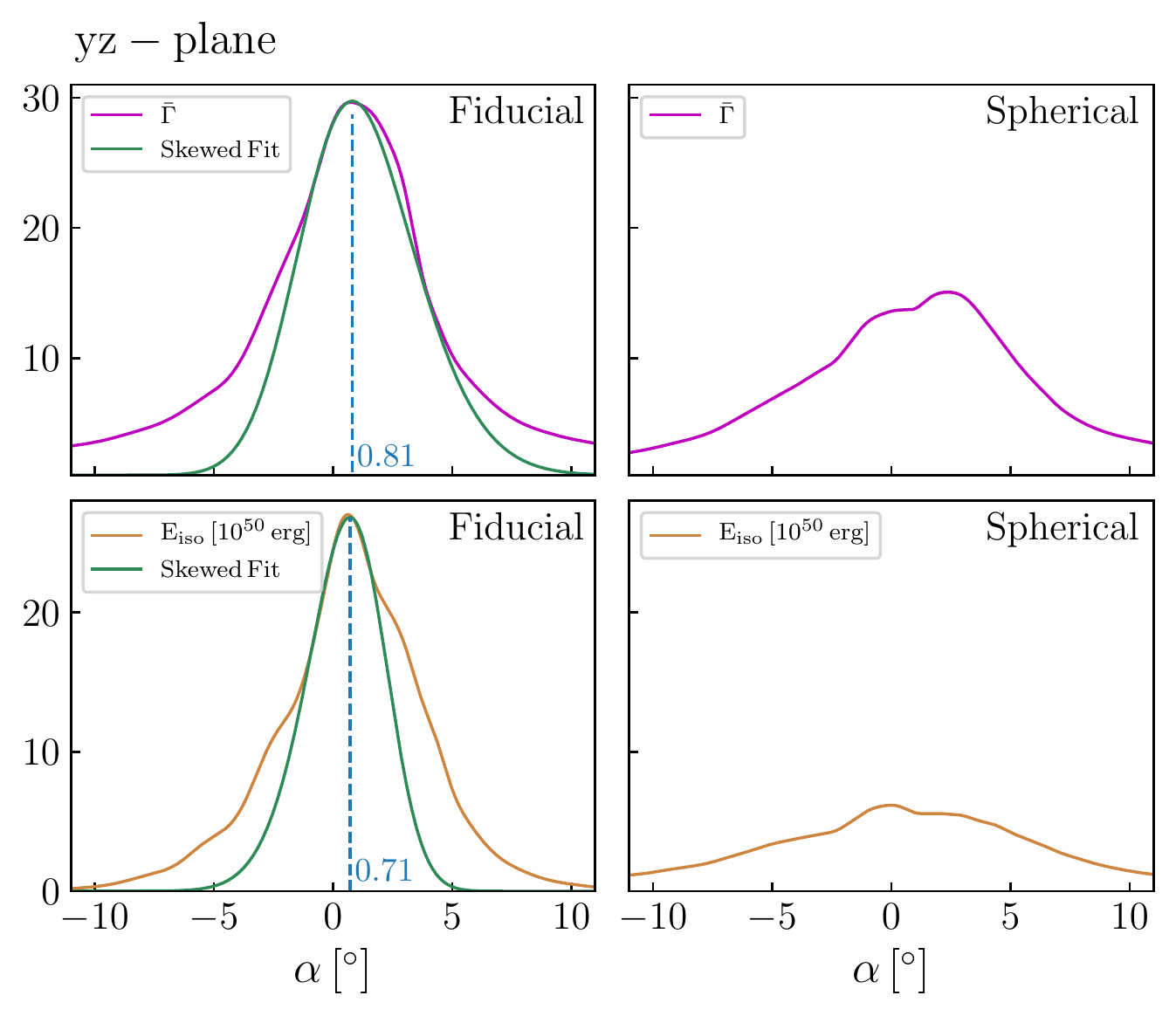}
   \caption{Same as Figure~\ref{GammaVSalpha_xy} for the $yz$-plane. In this case, the sign of $\alpha$ corresponds to the sign of $z$.} 
   \label{GammaVSalpha_yz}
\end{figure}
\begin{figure*}
   \includegraphics[width=1.6\columnwidth]{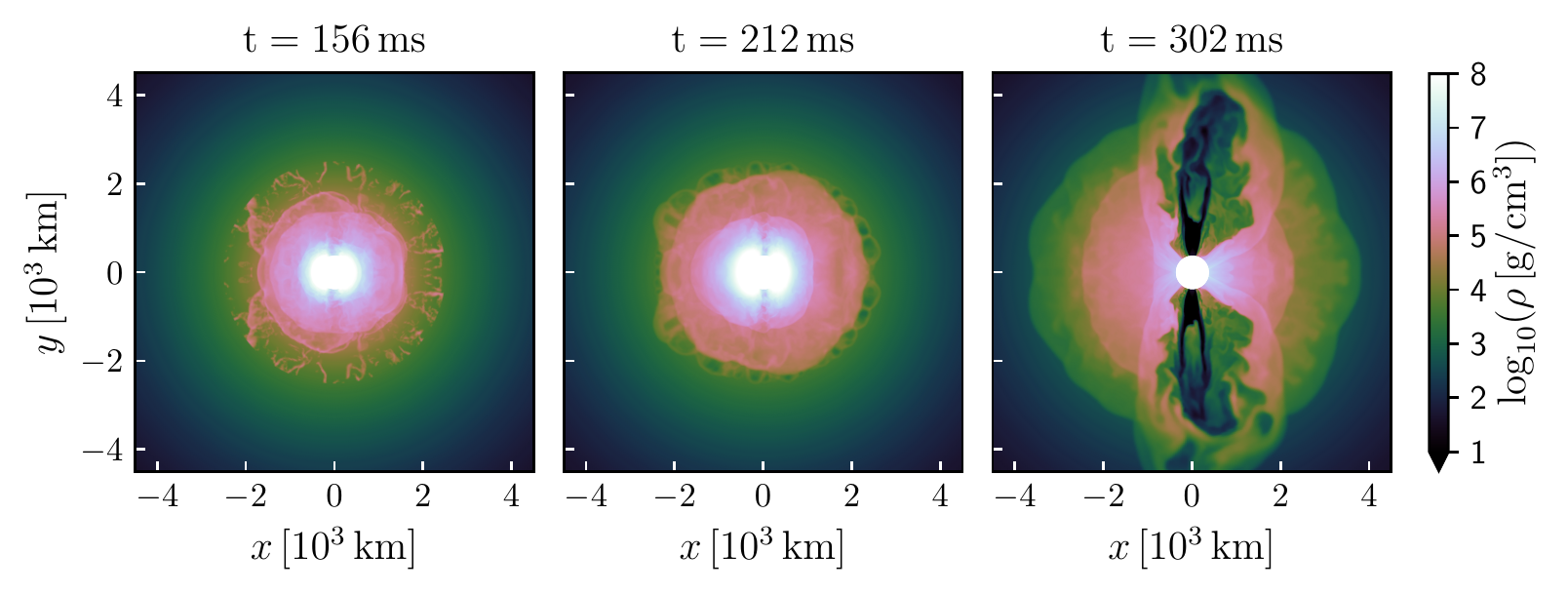}
   \caption{Same as Figure~\ref{rho_small} for the model with collapse at 201\,ms (and jet launching at 212\,ms).}
   \label{rho_small2}
\end{figure*}

Figures~\ref{large_scale} and \ref{gamma_max} (bottom row) show the simulation results in terms of rest-mass density, internal energy density, and Lorentz factor at the final time of 1012\,ms after merger.
To evaluate the effects of a simplified analytical and isotropic environment, we compare with the results of the previous simulation (without external forces, central row in Figs.~\ref{large_scale} and \ref{gamma_max}). The impact is substantial. The absence of a lower density funnel along the injection axis in the surrounding material makes it much harder for the jet to emerge, resulting in a breakout time delayed by $\simeq\!100\,$ms, a final maximum Lorentz factor around 30 (more than a factor of 2 lower and notably with a non-negligible north/south difference), and a less compact jet's head that has reached only $\simeq\!2.5\times10^5$\,km at its front.
On the other hand, after the initial breakout, the collimation at the base of the jet is more persistent and the jet's tail maintains a conical structure (Figure~\ref{large_scale}, bottom row). Finally, the overall jet structure is nearly axisymmetric and well aligned with the injection axis, differently from the cases with a non-isotropic initial environment (Figure~\ref{large_scale}).

To further illustrate the effects of a simplified isotropic environment on the emerging jet structure, the right panels of Figures~\ref{GammaVSalpha_xy} and \ref{GammaVSalpha_yz} report the angular profiles of radial-averaged Lorentz factor (top) and isotropic equivalent energy (bottom) at the jet's head and at 1012\,ms after merger, on $xy$- and $yz$-planes, respectively. 
We refer again to the jet's head, which is defined in this case by the radial range 
$r\!\in (18,25)\times10^4$\,km (see Figure~\ref{gamma_max}, bottom panel).
A direct comparison with the fiducial case (left panels of the same Figures) reveals not only much lower peak values, but also significantly smaller deviations from axisymmetry, i.e.~there are smaller differences between positive and negative $\alpha$ values on each plane, as well as between the two planes.

These results indicate that simplified analytical prescriptions for the surrounding environment, corresponding to what is often assumed, may substantially weigh on the jet dynamics and morphology when compared to the more realistic conditions obtained in BNS merger simulations.

\section{Collapse at 0.2\,seconds after merger}
\label{fiducial200}

In this Section, we discuss the results of a simulation with collapse time of the remnant NS set to $201\,$ms after merger. Unlike the fiducial case presented in Section~\ref{fiducial}, here the collapse time is not covered by the BNS merger simulation, which is limited to $156\,$ms. Therefore, in order to obtain the initial data for the following incipient jet evolution, we first need to continue or extrapolate the evolution from 156 to 201\,ms.
We treat such a case with a double purpose: (i) showing the feasibility of this kind of extrapolation and (ii) investigating the effect of a significantly different remnant NS lifetime (by a factor 2 in this case).

The data imported at the latest available time of the BNS merger simulation provide a different environment around the remnant NS. Figure~\ref{rho_small2} (left panel) shows in particular the rest-mass density at that time. Compared to 101\,ms post-merger (Figure~\ref{rho_small}), we observe a larger cloud of slowly expanding material resulting from the nearly isotropic baryon-loaded wind from the remnant NS. Radial motion is nearly homologous and radial velocity reaches a maximum of $\approx\!0.03\,c$ at $\approx\!1.5\times10^3\,$km. 
The inner and higher density region up to $\approx\!500\,$km remains rather unchanged and a lower density funnel along the y-axis is still present.

In order to evolve the system in PLUTO from 156 to 201\,ms post-merger, we first perform a 30\,ms test simulation from 126 to 156\,ms after merger following the prescriptions given at the beginning of Section~\ref{gravity}, in particular for the radial boundary conditions at the excision radius. 
The result, discussed in Appendix~\ref{extr}, shows a good match with the actual BNS merger simulation, giving us confidence to extrapolate the evolution at later times.
We then import the original data at 156\,ms after merger and use a prolongation of the same time-varying radial boundary conditions to evolve up to 201\,ms (i.e.~for 45\,ms, not much longer than the 30\,ms of the test).

\begin{figure}
   \includegraphics[width=\columnwidth]{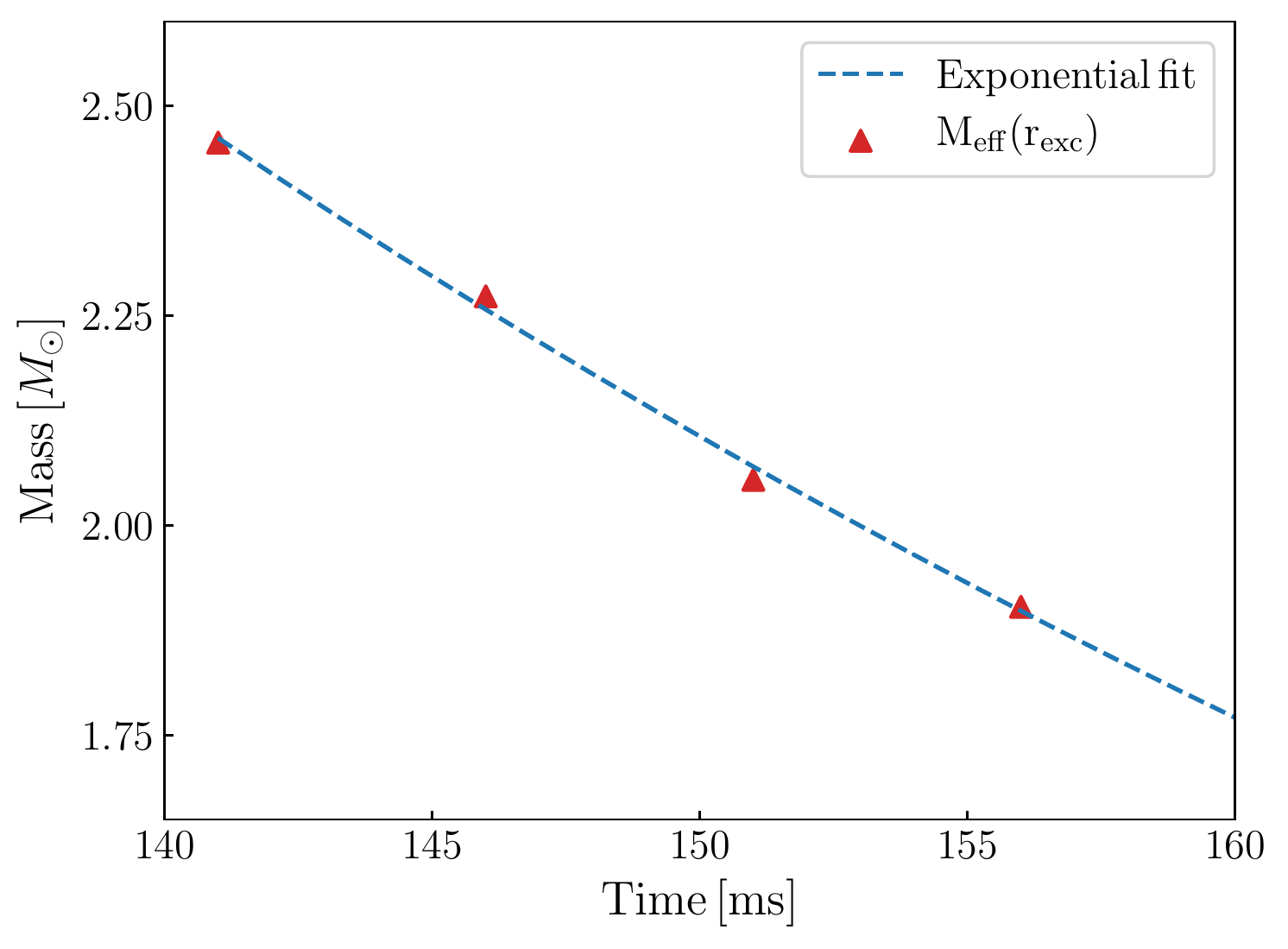}
   \caption{Evolution of the effective mass at the excision radius (380\,km) at times $\geq\!141\,$ms after merger. Red triangles are the values of $M_{\mathrm{eff}}(r_{\mathrm{exc}})$ extracted from the BNS merger simulation, while the dashed blue curve corresponds to an exponential fit (see text).}
   \label{Meff_decay}
\end{figure}

At $201\,$ms post-merger, the remnant NS is assumed to collapse and the following evolution consists, as in the fiducial case and the other cases discussed in Section~\ref{fiducial100}, of 11\,ms of post-collapse rearrangement and the subsequent jet injection.
The only missing ingredient to evolve the system after collapse is the initial value of the effective mass $M_\mathrm{eff}$ (see Section~\ref{gravity}), which cannot be retrieved in this case from the BNS merger simulation. 
As shown in Figure~\ref{Meff_decay}, the last part of the original simulation reveals a clear decreasing trend in the radial pressure gradients, which corresponds to a decreasing effective mass at the excision radius. An exponential decay with characteristic damping time of 58.8\,ms reproduces well the decreasing profile. 
Adopting the corresponding fitting function, we obtain the value of $M_\mathrm{eff}$ at the desired time (201\,ms after merger).

Also in this case, the simulation covers up to 900\,ms after the jet launching time, i.e.~up to $1112\,$ms after merger. 
Figure~\ref{rho_small2} shows the rest-mass density at the time of jet launching and around the time the jet itself breaks out of the surrounding environment (central and right panels, respectively).
We note that the total rest mass outside the excised region at the jet launching time ($\simeq\!9.4\times 10^{-2}\,M_\odot$) is a factor $\simeq\!3.6$ larger than in our fiducial model, due to the longer remnant NS lifetime. As a consequence, the incipent jet takes significantly longer to break out ($\simeq\!90$\,ms instead of $\simeq\!30$\,ms).

The effects of a longer NS lifetime and a later jet launching on the final jet structure (900\,ms after jet launching) can be appreciated in Figure~\ref{large_scale2}, where we show the rest-mass density and the Lorentz factor at large scales on the $xy$-plane.
Even though the jet injection parameters are the same as in the fiducial model, we observe important differences that are to be attributed to the different initial environment. 
The jet needs to drill through a more massive and extended cloud of material, spending more power to break out and reaching, after 900\,ms, a significantly smaller radial distance (jet's head front is located at $\simeq\!2.2\times10^5\,$km, to be compared with $\simeq\!2.7\times10^5\,$km of the fiducial model).
Also the maximum Lorentz factor at the jet's head is smaller ($\simeq\!22$ instead of $\simeq\!40$).
At the same time, the denser environment close to the excision surface maintains a higher degree of collimation and axisymmetry in the outflow.

For a given BNS merger model, the time interval between merger and jet launching is confirmed as a key parameter in determining the ultimate jet structure, which, in turn, shapes the corresponding radiative signatures. This offers good prospects for constraining such a time interval via the comparison with the observations (e.g., \citealt{Zhang2019,Gill2019,Lazzati2020,Beniamini2020}).

\section{Summary and conclusions}
\label{summary}

In this paper, we presented 3D special relativistic hydrodynamic simulations of incipient SGRB jets propagating through the baryon-polluted environment surrounding the remnant of a BNS merger. For the first time, we employ initial data for the environment obtained by directly importing the outcome (i.e.~density, pressure, and velocity distributions) of a general relativistic BNS merger simulation. 
This represents a first key step towards a consistent end-to-end description connecting the details of a specific BNS merger with the ultimate EM signatures associated with the breakout and propagation of an emerging SGRB jet. 

The simulations are performed with the PLUTO code, using rotated spherical coordinates (with polar axis lying on the BNS orbital plane) and logarithmic spacing along the radial direction.
A central sphere of $380\,$km radius is excised and suitable boundary conditions are adopted on the corresponding spherical surface. 
We also include the gravitational pull of the central object (with a mass of $\simeq\!2.596\,M_{\odot}$) and, after collapse, an extra (time and space dependent) radial acceleration term added to mimic a gradual fading away of the radial pressure gradient support close to the excision surface, as expected in a realistic post-collapse evolution.
The computational domain is initially filled with an artificial atmosphere with density and pressure scaling with distance as $\propto\!r^{-5}$ and zero velocity (also tested along with different choices of the power-law exponent). 
Finally, we adopt the Taub EOS. Being slightly different from the EOS used in the reference BNS merger simulation (at the low rest-mass densities of interest, i.e.~$<\!10^8$\,g/cm$^3$), small differences arise in the specific internal energy of the initial setup. However, dedicated benchmarks show that the influence on the final outcome is minor.

For the jet injection, we adopt the paradigm in which the central engine powering the relativistic outflow is an accreting BH-disk system, formed after the eventual collapse of a massive NS remnant.\footnote{In principle, our approach is also applicable to the case of a massive NS central engine. In this case, however, the formation of a jet might be difficult to justify unless directly covered (or at least strongly supported) by the BNS merger simulation itself.}
In this work, we assume that the collapse occurs at a chosen time after merger. This is the time at which we import data from the reference BNS merger simulation and start the evolution in PLUTO. 
The jet is launched after a short time window (set here to 11\,ms) from the collapse of the remnant NS, compatibly with the expected delay characterizing the formation of an incipient jet from a newly formed BH-disk system.

The incipient jet properties are the same in all our jet simulations. 
A top-hat outflow is continuously injected from the excision surface and within a half-opening angle of $10^{\circ}$ around the remnant spin axis (or orbital axis of the BNS). 
The initial luminosity, Lorentz factor, and specific enthalpy are $L_0\!=\!3\!\times\!10^{50}$\,erg/s, $\Gamma_0\!=\!3$, and $h_0\!=\!100$, respectively. 
An exponential time decay in luminosity is enforced with a characteristic timescale of 0.3\,s, which is consistent with the order of magnitude of typical accretion timescales of BH-disk systems formed in BNS mergers.

Our simulations probe two different collapse times, namely 101 and 201\,ms after merger. 
For our fiducial case (collapse at 101\,ms), we also repeat the simulation without the contribution of external forces (i.e.~gravity and the extra acceleration compensating for the missing radial pressure gradients after collapse) and in one case we also substitute the environment with an isotropic and homologously expanding one. \\

The main results of our study can be summarized as follows:
\begin{itemize}
    \item \textit{Realistic post-merger environment.}
    The density and velocity distributions of the material surrounding the merger remnant at the jet launching time, which depend on the details of the specific BNS system, can deviate significantly from the simplified isotropic and homologously expanding medium often considered in SGRB jet propagation studies. Comparing a reference model (with 112\,ms post-merger jet launching time and no external forces) with an equivalent one where the environment is substituted by the best-fitting isotropic and homologously expanding medium having the same total mass (see Section~\ref{fiducial100}), we find major differences in the outcome. The presence of a lower density funnel along the remnant spin axis allows the incipient jet to breakout more efficiently, retaining a higher energy. This, in turn, results in a much larger maximum Lorentz factor ($\Gamma \gtrsim\!70$ vs.~$\simeq\!30$, at 1012\,ms after merger) and a higher degree of collimation at the jet's head. Another relevant effect is caused by deviations from axisymmetry in the environment (e.g., due to the remnant recoil in unequal mass mergers), which make the final jet slightly misaligned with respect to the orbital axis and significantly less axisymmetric in structure.\footnote{This may also translate in uncertainties in GW-based Hubble constant estimates (e.g., \citealt{LVC-Hubble,Hotokezaka2019}).} \\
    This example poses a strong caveat for any model neglecting the anisotropy in matter distribution of the post-merger environment. Moreover, it shows that the final jet properties can be affected by other features that are typically not considered, such as deviations from axisymmetry or velocity distributions that are more complex than a simple homologous expansion.
\begin{figure}
   \includegraphics[width=\columnwidth]{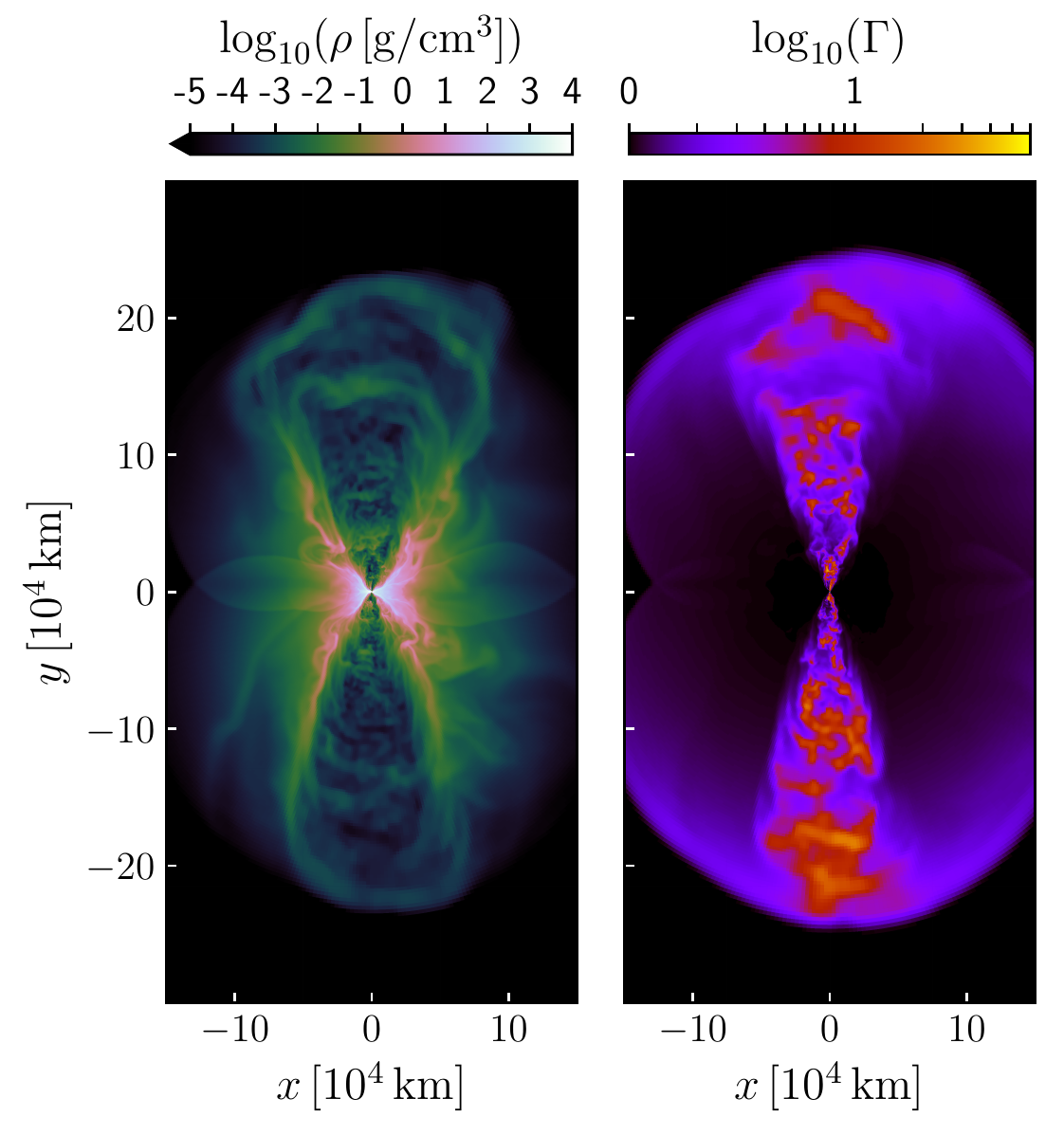}
   \caption{Meridional view of rest-mass density (left) and Lorentz factor (right) at $1112\,$ms after merger for the case with remnant NS collapse at $201$\,ms. The spatial and color scales used here are the same adopted in Figure~\ref{large_scale}.}
   \label{large_scale2}
\end{figure}

    \item \textit{Impact of gravity.} A proper description of the environment dynamics should take into account the gravitational pull of the central object, as also shown by the direct comparison with the BNS merger simulation results (e.g., Figure~\ref{1DExtrap_NEW}).
When gravity is included, the dynamics of the jet propagation is significantly affected. In particular, the surrounding material is allowed to fall back towards the central engine. The falling material directly encountered by the incipient jet acts as an obstacle, dissipating part of the jet energy into heat and turbulent motions. At the same time, accretion keeps reducing the overall mass of the environment, changing the lateral pressure balance between the jet and the surrounding material in favour of the former and leading to a significant widening of the jet's opening angle shortly above the injection radius. 
    The decreasing collimation ultimately results in a more compact jet's head followed by a much wider and slower outflow. 
    Finally, due to the gravitational pull acting on the environment material, the internal and kinetic energies of the emerging outflow only carry about half of the total injected energy. 
    Compared to the equivalent case with no gravitational pull, the maximum Lorentz factor achieved is much lower (almost a factor of 2 at 1012\,ms after merger), more energy is deposited in the cocoon, and there is no well defined jet's tail.\\
    In conclusion, the gradual and continuous accretion of the surrounding material, while being overlooked in most SGRB jet propagation models, can have a strong influence on the emerging jet properties.
    
    \item \textit{Fiducial model.} Our fiducial simulation, where the remnant NS collapses at 101\,ms after merger and the incipient jet is launched 11\,ms later, results in a final jet that has successfully emerged from the BNS merger environment.
    At 1012\,ms after merger, the internal-to-kinetic energy ratio of the outflow (for $r\!>\!3000$\,km) is about 15\%, indicating an advanced stage of conversion of heat into motion. The angular profiles of Lorentz factor and isotropic-equivalent energy at the jet's head reveal a central narrow core (of half-opening angle $\simeq\!3^{\circ}$ and $\simeq\!2^{\circ}$, respectively) surrounded by a wider and moderately relativistic outflow carrying a significant fraction of the total energy.
    These angular profiles also appear very different on the $xy$- and $yz$-planes, showing strong deviations from axisymmetry. Moreover, Lorentz factor and isotropic-equivalent energy peak along a direction that is slightly tilted with respect to the injection axis (by $0.7^{\circ}-0.9^{\circ}$). Such angular dependences cannot be reproduced via simple Gaussian or power-law functions. A skewed normal function can fit well the central core, but not the very asymmetric lateral wings. 
    This result suggests that employing simple functions to fit SGRB jet angular structures (as revealed, e.g., by afterglow observations) may require some caution. 
    
    \item \textit{Dependence on the jet launching time.} When considering a jet launching time of 212\,ms post-merger (almost double with respect to the fiducial case), the very same incipient jet has to drill through a significantly more massive environment (factor $\simeq\!3.6$). As a consequence, more energy is dissipated into the surrounding material, it takes longer to break out ($\simeq\!90$\,ms vs.~$\simeq\!30$\,ms), and the maximum Lorentz factor reached is lower ($\simeq\!22$ vs.~$\simeq\!40$ at 900\,ms after jet launching). On the other hand, the more expanded and massive environment provides a more efficient collimation at the base of the jet, also resulting in a more axisymmetric final structure. The comparison with the fiducial case confirms that the time interval between merger and jet launching can have a strong influence, thus offering the opportunity to tightly constrain such a key parameter via observations. \\
While we consider here jet launching times of up to $\approx\!200$\,ms after merger, significantly longer delays are possible. In the case of GRB\,170817A, for instance, there is an ample range of delays favoured by different authors, going from a few hundred ms (e.g., \citealt{Zhang2019,Lazzati2020}) to order $\sim\!1$\,s (e.g., \citealt{Gill2019,Murguia2021}).
Within our setup, considering later jet launching times would require longer BNS merger simulations, beyond our current reach, and/or more extended extrapolations. The latter, to be considered reliable, would likely demand a more refined scheme (compared to what is done here) and in-depth testing against BNS merger simulation results.
A possible consequence of a later jet launching time could be that, due to the slower rate of change in the environment mass, the final outcome will depend less on the time delay itself. This represents an interesting issue open for investigation.
\end{itemize}

The main aim of this work is to introduce a new approach to address the problem of SGRB jet propagation in BNS merger environments, showing the potential advantages of employing the outcome of BNS merger simulations as initial data.
The prescriptions and assumption adopted here, while attempting to offer a more realistic description of some aspects of the system dynamics, leave plenty of room for further improvement. 
In particular, we do not include magnetic fields, which are a key ingredient in SGRB jet production and evolution. 
Furthermore, the incipient jet is introduced by hand and not produced self-consistently in the BNS merger simulation. 
Overcoming the above limitations (among others) should represent a priority in future studies.

\section*{Acknowledgements}

We thank the anonymous referee for very helpful remarks and constructive comments on the manuscript.
We also thank Om Sharan Salafia and Stefano Ascenzi for useful discussions. 
J.V.K. kindly acknowledges the CARIPARO Foundation for funding his PhD fellowship within the PhD School in Physics at the University of Padova.
All the simulations were performed on GALILEO and MARCONI machines at CINECA. In particular, we acknowledge CINECA for the availability of high performance computing resources and support through awards under the ISCRA and the MoU INAF-CINECA initiatives (Grants \texttt{IsB18\_BlueKN, IsB21\_SPRITZ, INA20\_C6A49, INA20\_C7A58}) and through a CINECA-INFN agreement, providing the allocations \texttt{INF20\_teongrav} and \texttt{INF21\_teongrav}.

\section*{Data Availability}

The data underlying this article will be shared on reasonable request to the corresponding authors.



\bibliographystyle{mnras}
\bibliography{refs} 


\appendix

\section{Dependence on atmosphere prescription}
\label{atmo}
\begin{figure}
   \includegraphics[width=\columnwidth]{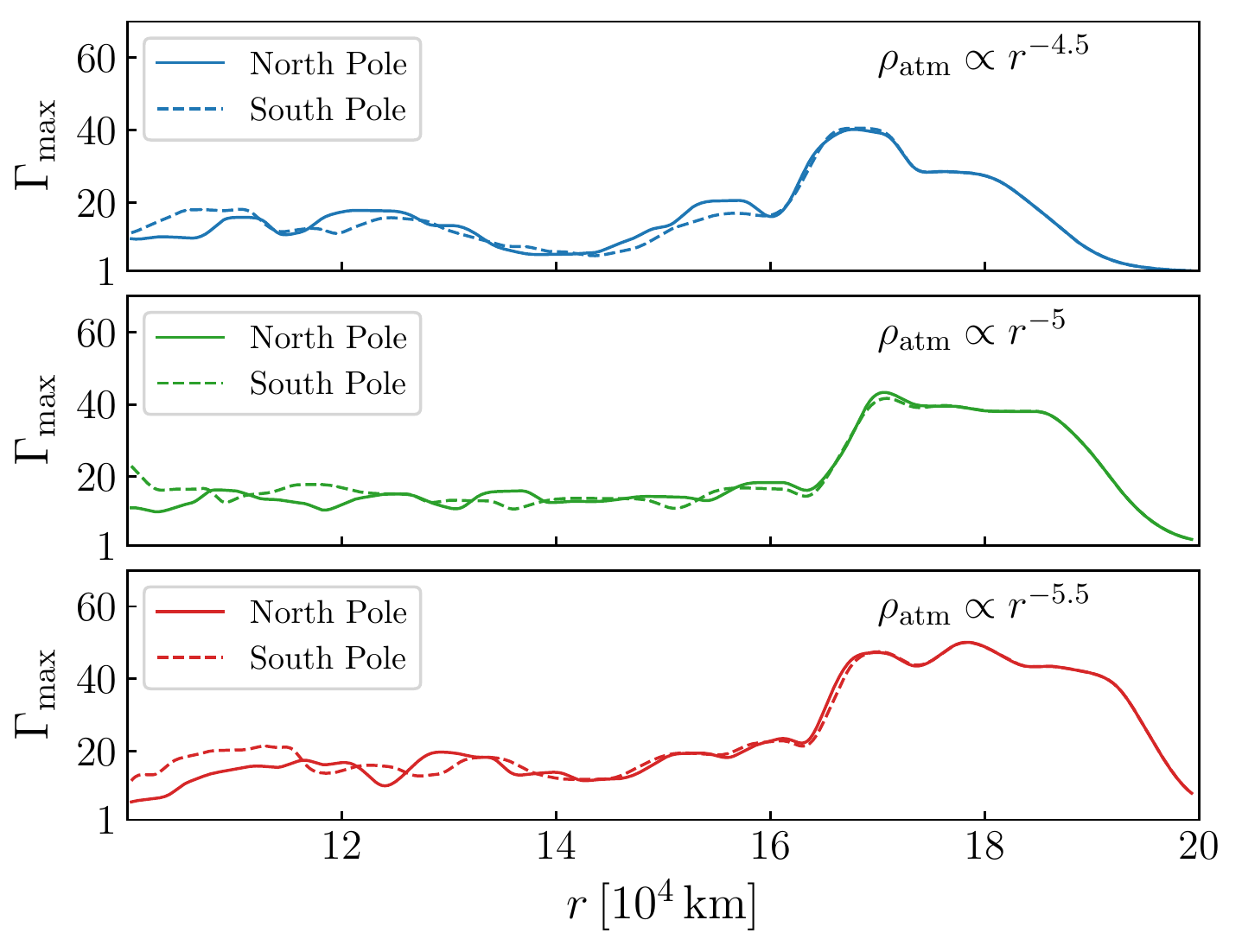}
   \caption{Same as Figure~\ref{large_scale} for our fiducial model with different power-law exponents for the radial profile of density and pressure in the atmosphere, at 762\,ms after merger.}
   \label{gamma_atmo}
\end{figure}

When setting up the initial conditions for our simulations, we add an artificial atmosphere characterized by radial profiles of density and pressure decaying as $r^{-a}$ (see Sect.~\ref{import}).
Here, we investigate the potential impact of such an atmosphere on the jet propagation by considering our fiducial model (Sect.~\ref{fiducial}) with three different values of the power-law exponent $a$, namely $a\!=\!4.5$, 5, and 5.5. In this case, we run the simulations up to 762\,ms after merger.

Since the initial atmosphere can represent an obstacle for the emerging jet, the corresponding braking effect is mostly visible when looking at the distance reached by the jet's head at the latest available time.
Figure~\ref{gamma_atmo} shows the radial profile of the maximum Lorentz factor (i.e.~the maximum reached at given radial distance) at 762\,ms after merger. Going from $a\!=\!4.5$ to 5, we notice differences in the jet's head profile and distance reached. From 5 to 5.5 differences are significantly reduced.
We conclude that the power-law exponent should be set to a value of at least 5 (as in our fiducial case).

In order to present quantitatively accurate results, future studies should explore even larger power-law exponents and include any residual atmospheric effects in the error budget.

\section{Impact of EOS change}
\label{EOS}
\begin{figure}
   \includegraphics[width=\columnwidth]{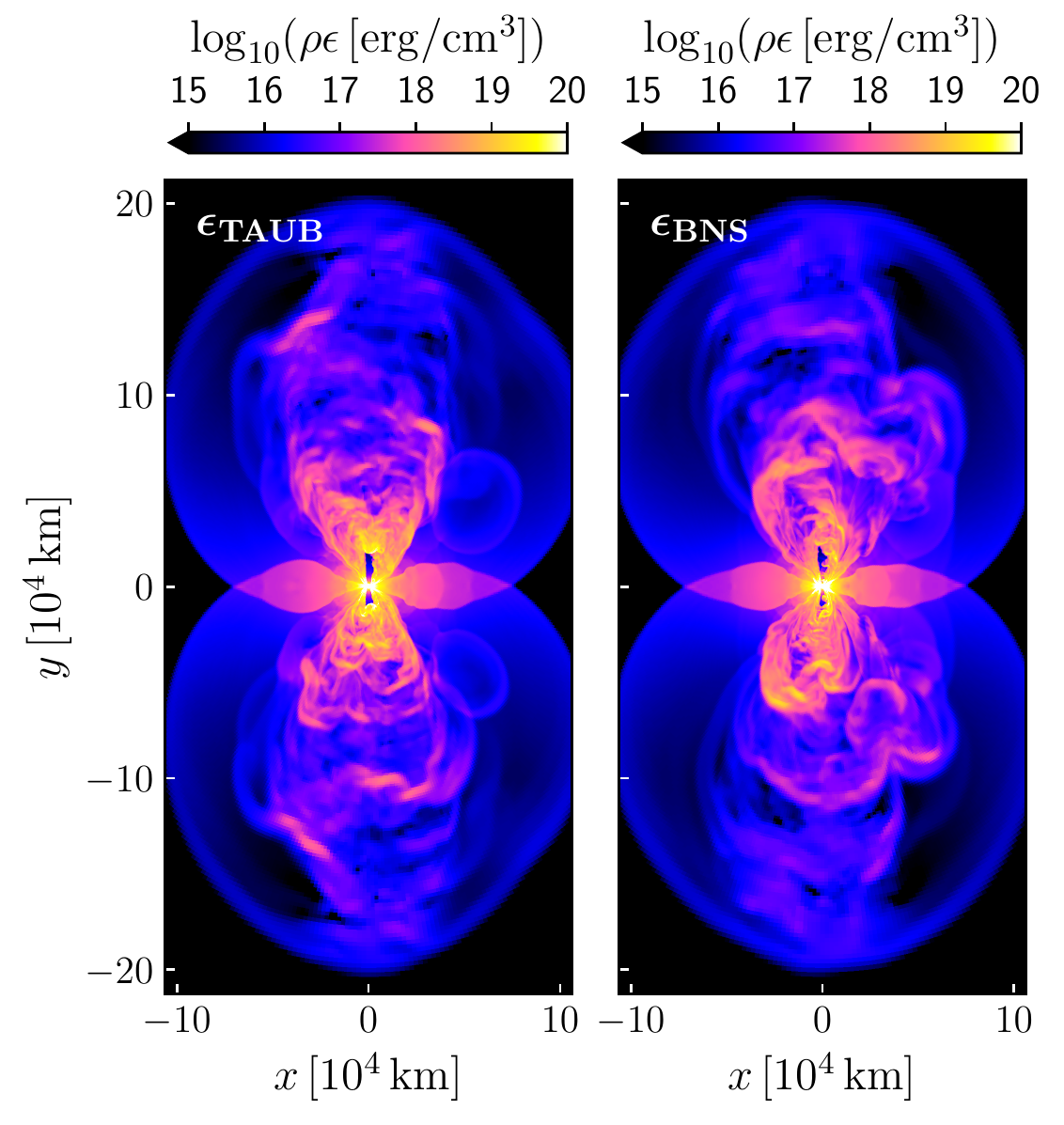}
   \caption{Meridional view of internal energy density at 762\,ms after merger. Left panel refers to our fiducial model, where the initial rest-mass density and pressure are directly imported from the reference BNS merger simulation, while the specific internal energy $\epsilon$ is derived via the Taub EOS. For the case on the right, the imported quantities are instead rest-mass density and $\epsilon$, while the quantity derived via the Taub EOS is the pressure.}
   \label{large_scale_eos}
\end{figure}
\begin{figure}
   \includegraphics[width=\columnwidth]{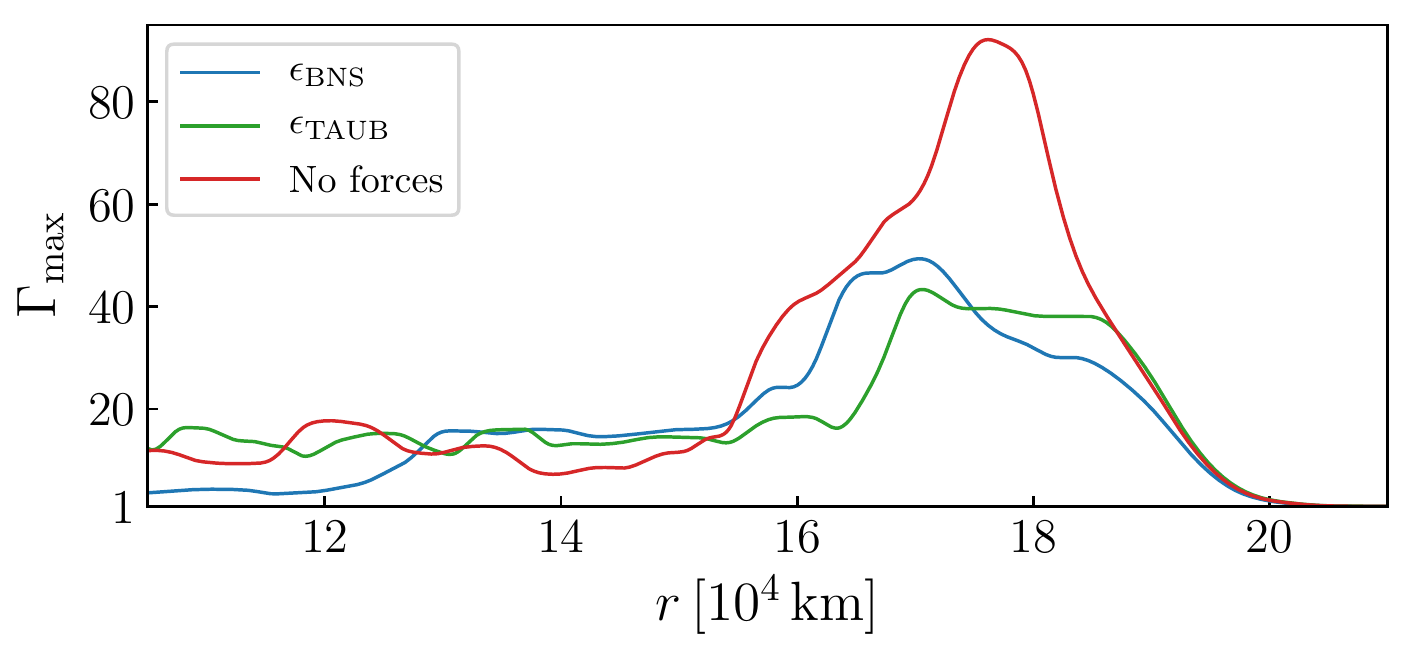}
   \caption{Radial profiles (north side only) of maximum Lorentz factor at 762\,ms after merger. The three profiles refer to our fiducial model ($\epsilon_\mathrm{\,TAUB}$), the one where specific internal energy is directly imported ($\epsilon_\mathrm{\,BNS}$; see text), and the one where no external forces are included.}
   \label{gamma_eos}
\end{figure}

As pointed out in Section~\ref{import}, the Taub EOS employed in our simulations does not exactly match the EOS used in the BNS merger simulation (at the low densities of interest). This leads to a slight discrepancy in the initial internal energy density, which is recomputed via the Taub EOS from the imported rest-mass density and pressure data values. 
In order to obtain an indicative measure of the effects on our results, we compare here two simulations: the first one corresponds to our fiducial model (Sect.~\ref{fiducial}), while the second one has an initial setup with imported rest-mass density and specific internal energy (and the pressure is recomputed via the Taub EOS).
The latter simulation covers up to 762\,ms after merger.

At the level of initial data, the two cases differ by $\simeq\!7\%$ in (total) internal energy and by $\simeq\!6\%$ in the sum of kinetic and internal energy. As the system evolves, a significant portion of material falls-back across the excision surface (due to the gravitational pull; Sect.~\ref{gravity}) and, at the same time, the incipient jet brings in a significant amount of additional internal and kinetic energy. The combination of the two effects makes the initial discrepancy in energy less and less important as the evolution proceeds. Indeed, at 762\,ms post-merger, the sum of kinetic and internal energy only differs by $\simeq\!2\%$ and such a difference keeps decreasing.
Figure~\ref{large_scale_eos} shows the comparison for the internal energy density at 762\,ms, where no significant differences are present.

As expected from the above considerations, the variation in terms of emerging jet properties is rather limited. 
In Figure~\ref{gamma_eos}, we report the radial profile of the Lorentz factor (maximum value at each radial distance) at 762\,ms.
The radial location and profile of the jet's head is found to be rather similar and the difference in the overall maximum is $\sim\!10\%$.
For comparison, we also show the case where no external forces are considered (with specific internal energy recomputed via the Taub EOS; see Sect.~\ref{forces}), for which the difference in the overall maximum is much larger (factor $\simeq\!2$).

For the purposes of the present work, the effects due to the mismatch in the initial internal energy are acceptable.
On the other hand, a fully consistent EOS would eliminate such a potential source of error. This could be obtained by adapting the EOS employed in the BNS merger simulation so that it reproduces the Taub EOS below rest-mass densities of $10^8$\,g/cm$^3$.

\section{Resolution study}
\label{res}
\begin{figure*}
   \includegraphics[width=1.9\columnwidth]{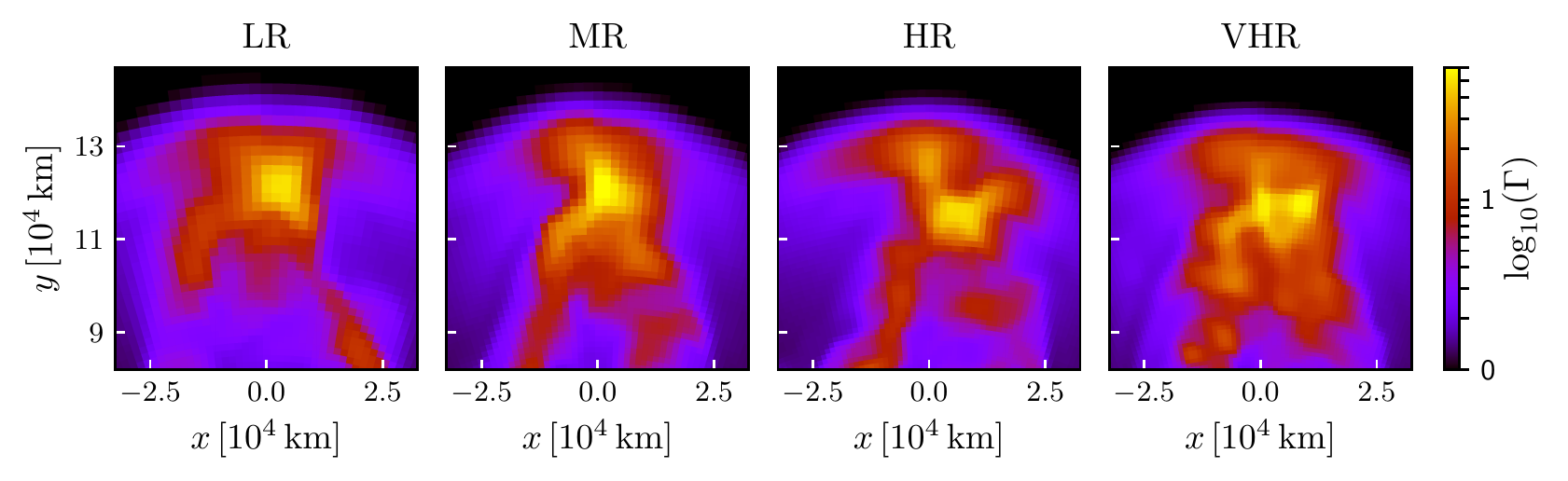}
   \caption{Meridional view of the Lorentz factor at the jet's head (north side only) computed at $572\,$ms after merger. The four panels show the results at four different resolutions (left to right): LR, MR, HR and VHR (see text for details).}
   \label{large_scale_res}
\end{figure*}
\begin{figure}
   \includegraphics[width=\columnwidth]{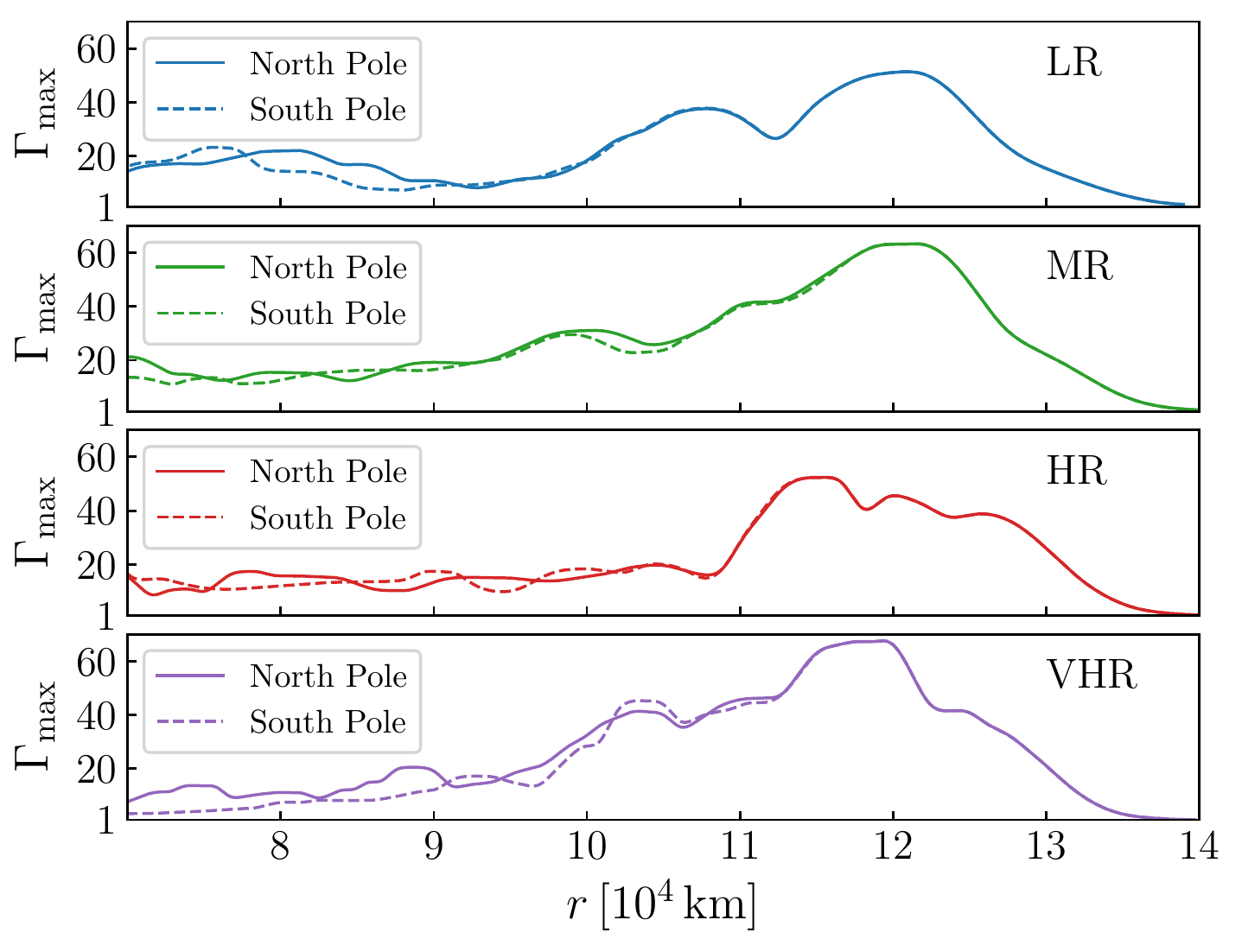}
   \caption{Radial profiles of maximum Lorentz factor at $572\,$ms after merger (maximum value achieved at each radial distance). North and south profiles are shown for the four different resolutions (top to bottom): LR, MR, HR and VHR (see text for details).}
   \label{Gmax_res}
\end{figure}

To evaluate how our results depend on resolution, we performed simulations of our fiducial model (Sect. \ref{fiducial}) with four different settings:
\begin{itemize}[leftmargin=+.3cm]
\item Low Resolution (LR): $540\times180\times360$ points along $r$, $\theta$, and $\phi$, respectively\,;
\item Medium Resolution (MR): $648\times216\times432$ points\,; 
\item High Resolution (HR): $756\times252\times504$ points\,; 
\item Very High Resolution (VHR): $864\times288\times576$ points\,,
\end{itemize}
where HR is our fiducial resolution.
Below, we compare the outcome at $572\,$ms after merger. In particular, we focus on the Lorentz factor distribution at that time.

In Figure~\ref{large_scale_res}, we show the meridional view of the Lorentz factor, zooming in the region of the jet's head (north side only), with resolution increasing from left to right.
In Figure~\ref{Gmax_res}, we report the radial profile of the maximum Lorentz factor computed for the different resolutions (increasing from top to bottom). 
From both Figures, we can appreciate the gradual appearance of finer spatial modulations as the resolution increases. Significant differences are still present between the two highest resolutions, i.e.~HR and VHR, which indicates that we are not yet in a regime of convergence. 
While a precise assessment of numerical errors is beyond our present scope, future studies presenting quantitative results will thus require higher resolutions (corresponding to our VHR or higher).

\section{Extrapolation test}
\label{extr}
\begin{figure}
   \includegraphics[width=0.96\columnwidth]{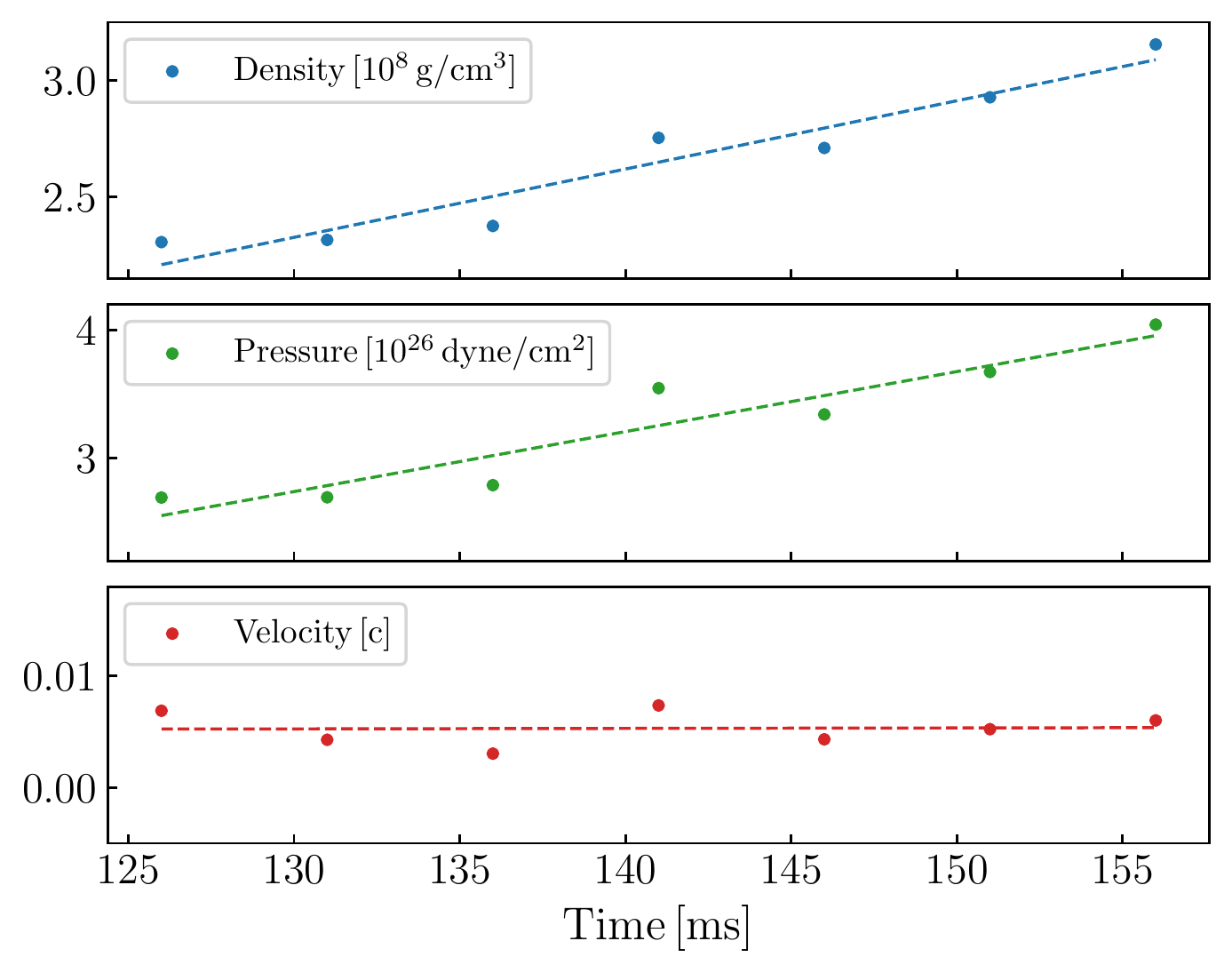}
   \caption{Time evolution of the angle-averaged rest-mass density, pressure, and radial velocity (top to bottom) at the excision radius, from 126 to 156\,ms after merger. Filled circles correspond to values extracted from the original BNS merger simulation. Dashed lines are the result of linear fits to such data.}
   \label{meandata_126}
\end{figure}
\begin{figure}
   \includegraphics[width=\columnwidth]{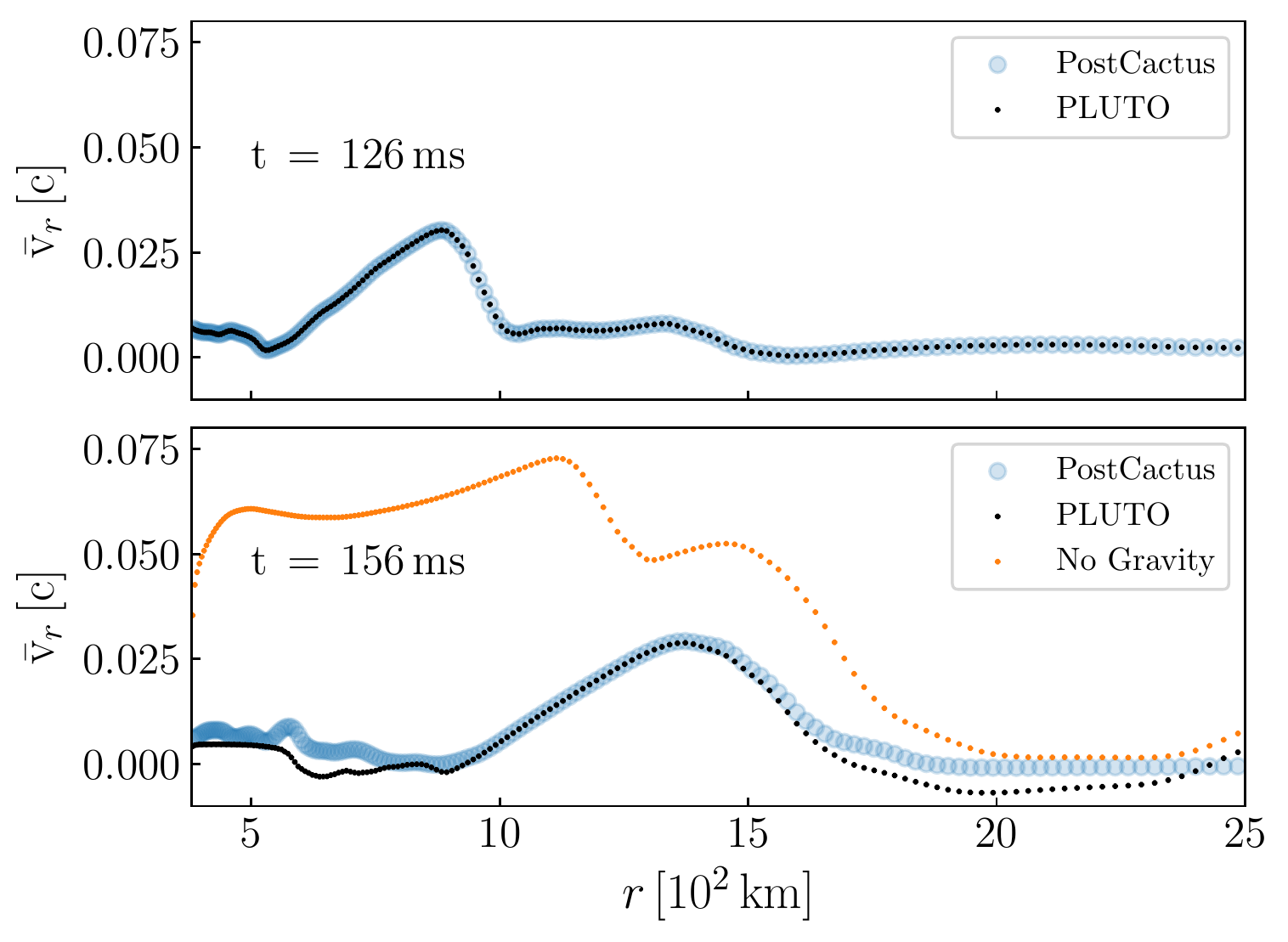}
   \caption{Radial profiles of the angle-averaged radial velocity at $126\,$ms and $156\,$ms after merger (upper and lower panels, respectively). The blue dots are obtained by importing data from our reference BNS merger simulation at the given time (Section~\ref{import}). The black dots correspond instead to the result of our test PLUTO simulation, which starts from the same initial data at 126\,ms and then evolves the system according to the adopted prescriptions (see text for details). Finally, the orange dots correspond the same PLUTO simulation where we neglect the contribution of gravity.}
   \label{1DExtrap_NEW}
\end{figure}
\begin{figure}
   \includegraphics[width=\columnwidth]{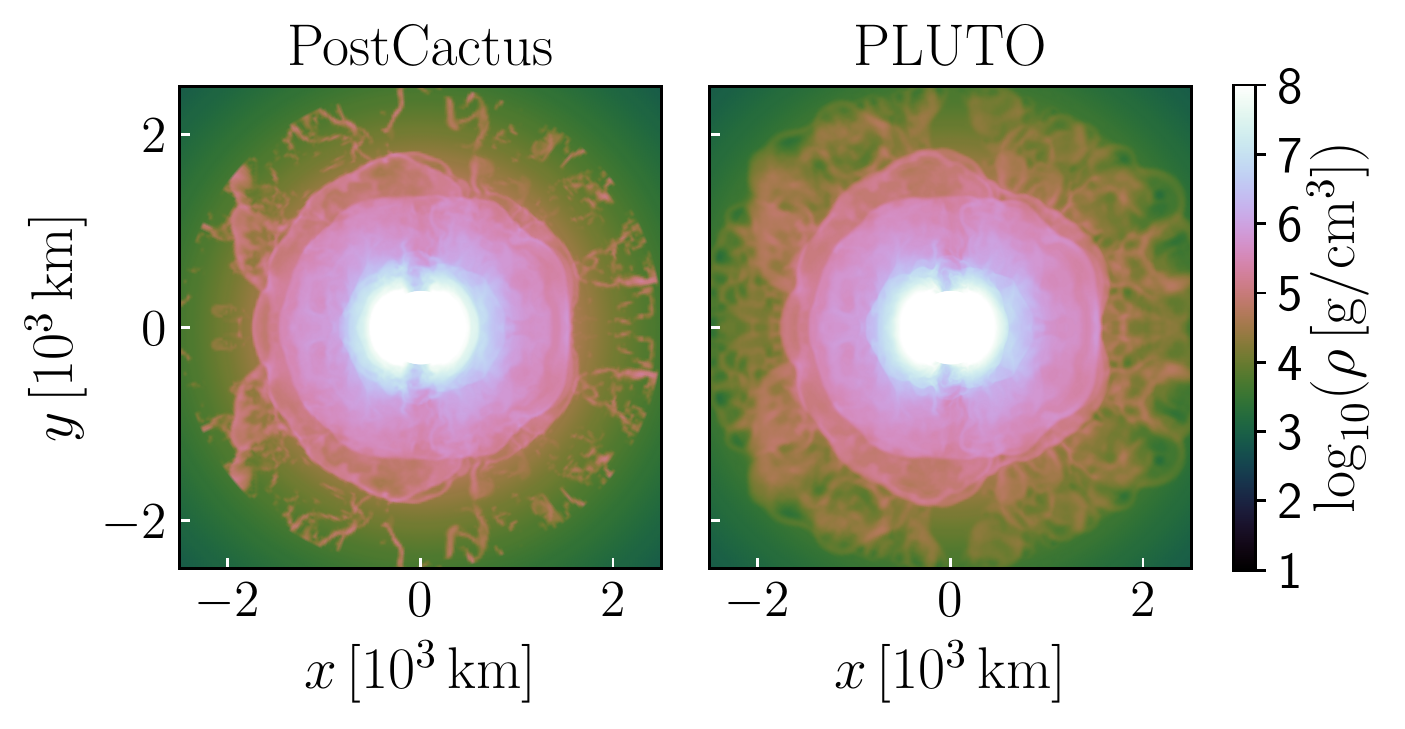}
   \caption{Meridional view of the rest-mass density at $156\,$ms after merger for the same two cases shown in Figure~\ref{1DExtrap_NEW}, i.e.~data imported from our reference BNS merger simulation (left) and the result of the evolution in PLUTO starting from data imported at 126\,ms (right). The white circle of 380\,km radius corresponds to the excised region (as in Figs.~\ref{fig0} and \ref{rho_small}).}
   \label{2Drho_126_NEW}
\end{figure}

We present here the results of a PLUTO simulation where we import data from the BNS merger simulation at 126\,ms after merger and evolve the system up to 156\,ms with the prescriptions given in the first part of Section~\ref{gravity} (with no collapse to a BH, nor jet injection).
Namely, we first obtain linear fits to the original time evolution of the angle-averaged rest-mass density, pressure, and radial velocity at the excision radius (Figure~\ref{meandata_126}). Then, we impose inner radial boundary conditions such that the initial distributions of rest-mass density, pressure, and 3-velocity persist on the excision sphere, but each quantity evolves in time according to the obtained linear trend, i.e. 
\begin{align*}
&\rho(r_\mathrm{exc},\theta,\phi,t) = \rho(r_\mathrm{exc},\theta,\phi,t_\mathrm{in}) \,\times\, F_{\rho}(t) \, , \\ 
&P(r_\mathrm{exc},\theta,\phi,t) = P(r_\mathrm{exc},\theta,\phi,t_\mathrm{in}) \,\times\, F_{P}(t) \, , \\
&\Vec{\mathrm{v}}(r_\mathrm{exc},\theta,\phi,t) = \Vec{\mathrm{v}}(r_\mathrm{exc},\theta,\phi,t_\mathrm{in}) \,\times\, F_{\mathrm{v}}(t) \, , 
\end{align*}
where $t_\mathrm{in}\!=\!126$\,ms after merger and
\begin{align*}
&F_{\rho}(t) = 0.96 + 0.013 \times (t-t_\mathrm{in})\mathrm{[ms]} \, , \\ 
&F_{P}(t) = 0.95 + 0.017 \times (t-t_\mathrm{in})\mathrm{[ms]} \, , \\
&F_{\mathrm{v}}(t) = 0.76 + 0.00061 \times (t-t_\mathrm{in})\mathrm{[ms]} \, .
\end{align*}
We note that density and pressure show a significant increase, while radial velocity remains essentially constant over time (Figure~\ref{meandata_126}).

Figure~\ref{1DExtrap_NEW} shows the radial profiles of the angle-averaged radial velocities at 126 and 156\,ms, for both the test PLUTO simulation and the original BNS merger simulation.
Starting from the very same profile at 126\,ms, the PLUTO simulation with the chosen prescription is able to nicely reproduce the final profile at 156\,ms (also close to the excision radius), with only a slight discrepancy around $600\,$km.
In the Figure, we also report the result obtained by eliminating the gravitational pull. The large discrepancy demonstrates how gravity is fundamental to reproduce the correct behaviour.

In Figure~\ref{2Drho_126_NEW}, we compare the rest-mass density in the $xy$-plane at 156\,ms resulting from the original BNS merger simulation and the PLUTO simulation. The correspondence is excellent. We also note that the lower density funnel along the $y$-axis is preserved.

In conclusion, the test is successful and suggests that our relatively simple prescription could also be used to continue or extrapolate in PLUTO the post-merger evolution beyond the time reached by the original BNS merger simulation, at least in this phase ($\gtrsim\!120$\,ms after merger).


\bsp	
\label{lastpage}
\end{document}